\documentclass[a4paper,11pt]{article}
\usepackage{jheppub}
\usepackage{multirow}
\usepackage{graphicx}
\usepackage{color}

\usepackage{atlasphysics}
\usepackage{mathptmx}
\usepackage{helvet}
\usepackage{epsfig}
\usepackage{subfigure}
\usepackage{amsmath}
\usepackage{preprintcover}

\PreprintCoverPaperTitle{Measurement of the top quark charge in \textit{\textbf{pp}}   collisions at $\bf{\sqrt{s}}$\ =\ 7\,TeV with the ATLAS detector}

\PreprintIdNumber{CERN-PH-EP-2013-056}  

\PreprintCoverAbstract{A measurement of the top quark electric charge is carried out in the ATLAS experiment at the Large Hadron Collider using 2.05 fb$^{-1}$ of data at a centre-of-mass energy of 7 TeV. 
In units of the elementary electric charge, the top quark charge is determined to be 0.64 $\pm $ 0.02 (stat.) $\pm $ 0.08 (syst.) from the charges of the top quark decay 
products in  single lepton $t\bar{t}$ candidate events.
This excludes  
models that propose a heavy quark of electric charge --4/3, instead of the Standard Model top quark, with a significance of more than 8$\sigma $.}
\PreprintJournalName{JHEP}  

\title{Measurement of the top quark charge in \textit{\textbf{pp}} collisions at $\bf{\sqrt{s}}$ = 7\,TeV with the ATLAS detector}

\author{The ATLAS Collaboration}
  
\newcommand{\SM}{Standard Model}

\def\TeVc{\ifmmode {\mathrm{\ Te\kern -0.1em V}}\else
                   {\textrm{Te\kern -0.1em V}}\fi}%
\def\GeVc{\ifmmode {\mathrm{\ Ge\kern -0.1em V}}\else
                   {\textrm{Ge\kern -0.1em V}}\fi}%

\def\TeVcc{\ifmmode {\mathrm{\ Te\kern -0.1em V}}\else
                   {\textrm{Te\kern -0.1em V}}\fi}%
\def\GeVcc{\ifmmode {\mathrm{\ Ge\kern -0.1em V}}\else
                   {\textrm{Ge\kern -0.1em V}}\fi}%

\abstract{A measurement of the top quark electric charge is carried out in the ATLAS experiment at the Large Hadron Collider using 2.05 fb$^{-1}$ of data at a centre-of-mass energy of 7 TeV. 
In units of the elementary electric charge, the top quark charge is determined to be 0.64 $\pm $ 0.02 (stat.) $\pm $ 0.08 (syst.) from the charges of the top quark decay 
products in  single lepton $t\bar{t}$ candidate events.
This excludes  
models that propose a heavy quark of electric charge --4/3, instead of the Standard Model top quark, with a significance of more than 8$\sigma $.}

\begin{document}
\maketitle
\flushbottom

\section{Introduction}
\label{intro}

\indent It is generally accepted that the particle discovered at Fermilab in 1995 \citep{abe95,d095} is the Standard Model (SM) top quark. However, a few years after the discovery 
a theoretical model appeared proposing an ``exotic'' quark of charge --4/3 and  mass $\approx$ 170 GeV as an alternative to the SM top quark at this mass value \citep{chang00}.  
Though this model has already been experimentally excluded a precise measurement of the top quark charge is important as it is one of the basic top quark properties.
A strong preference for the SM top quark with electric charge of $+2/3$ (in units of the electron charge magnitude)
was reported  by the D0 and CDF collaborations \citep{D0_topQ, CDF_topQ_2} but without the ultimate 5$\sigma $ exclusion of a possible exotic quark with charge of --4/3. The CDF and D0 exclusion limits are 95\%\footnote {The CDF collaboration has recently submitted an update of their analysis for publication, which results in a limit of 99\% \citep{cdfnew}.} and 92\%, respectively.
Therefore, it is still important to carry out a more precise measurement to definitively resolve this question with more than 5$\sigma $ confidence level. Due to the excellent ATLAS detector performance, the analysis presented here not only demonstrates that the particle presently denoted by ``top quark'' is really the SM top quark decaying into a $b$-quark and a $\Wboson^+$ boson, but also allows for a direct measurement of its electric charge with a significantly improved precision. 
 Moreover, from an experimental point of view it is interesting to demonstrate the high flavour tagging performance of the ATLAS experiment, i.e. its capability to distinguish between jets initiated by quarks and anti-quarks  used in this study to find the correct $Wb$ pairing in the   $W^+W^-b\bar{b}$ system of the assumed \ttbar\ final state.
 
\indent 
The dominant decay channel of the top quark is to a $b$-quark through the charged weak current:
$t \rightarrow W^{+}b$ $(\bar t \rightarrow W^{-}\bar b)$.
The measurement of the top quark charge requires the charges of both the \Wboson\ boson and the $b$-quark to be determined. While the former can be determined through W's leptonic decay, the $b$-quark charge is not directly measurable
due to quark confinement in hadrons. However, it is possible to establish a correlation between the charge of
the $b$-quark and the charges of the collimated hadrons from the $b$-quark hadronization that form a $b$-jet.
Within this approach,  the charge can be determined using the lepton$\,+\,$jets ($\ttbar\  \rightarrow \ell^{\pm}\nu jj\bbbar$) 
or the dilepton ($\ttbar\ \rightarrow \ell^{+}\nu \ell^{-}\bar \nu \bbbar$) channel. 
This paper presents the results of a top quark charge analysis based on the charges of the hadrons associated with the jet originating from a $b$-quark ($b$-jet) using the statistically more significant lepton$\,+\,$jets  channel. 

\section{The ATLAS detector}
\label{sec:atlas_det}

The ATLAS detector is a multi-purpose particle physics apparatus operating at the beam interaction point IP1 of the Large Hadron Collider (LHC). A complete description is provided in ref.~\citep{AtlDet2}.  
ATLAS uses a right-handed coordinate system with its origin at the centre of the detector (the nominal interaction point) and the $z$-axis along the beam pipe. The $x$-axis
points to the centre of the LHC ring, and the $y$-axis points upward.

The innermost part is an inner tracking detector (ID) comprising a silicon pixel detector, a silicon microstrip detector, and a transition radiation tracker. 
The inner detector covers the pseudorapidity\footnote{The pseudorapidity is defined in terms of the polar angle with respect to the beam axis, $\theta$,  as $\eta = -\ln(\tan(\theta /2))$.} range $\mid \eta \mid  <$~2.5 and is surrounded by a thin superconducting solenoid
providing a 2~T axial magnetic field, and by liquid-argon (LAr) electromagnetic sampling calorimeters with high granularity.
An iron/scintillator tile calorimeter provides hadronic energy measurements in the central pseudorapidity range ($\mid \eta \mid  < $~1.7). The end-cap and forward regions are instrumented with LAr sampling calorimeters for electromagnetic (EM) and hadronic energy measurements up to $\mid \eta \mid   =$ 4.9. The calorimeter system is surrounded by a muon spectrometer incorporating 
three superconducting toroid magnet assemblies, providing a toroidal magnetic field with bending power between 2.0 Tm and 7.5 Tm,  and a pseudorapidity coverage of $\mid \eta \mid <$~2.7.

\section{Data and simulation samples} \label{sec:data_mc}

\indent This analysis uses the proton--proton collision data collected
by the ATLAS experiment from March to August 2011 at a centre-of-mass energy of
$\sqrt{s}$ = 7~\TeV\, corresponding to an integrated luminosity of 2.05 $\pm $
0.04~\ifb \cite{Aad:2013lumi}. The data for the top quark charge study were collected using a single-muon and a single-electron trigger (see details in section \ref{sec:select}). In this analysis we also use the dijet data sample collected using the combined muon-jet trigger which requires
a reconstructed muon matched to a 10 GeV jet in the calorimeter.

Simulated event samples are used to estimate both the signal selection
efficiency and some of the background contributions and also to calibrate the $b$-jet charge measurement. 
The response of the ATLAS detector is simulated using {\sc Geant}4~\cite{Geant4_2003} and the resulting events are reconstructed by the same software~\cite{atlas_sim} used for data.

The MC@NLO Monte Carlo (MC) generator  v3.41, based on the next-to-leading-order (NLO) matrix elements \cite{MCatNLO,Frixione:2003ei} with CTEQ6.6 
\cite{PhysRevD.78.013004} parton distribution functions (PDFs), is used for the parton-level hard scattering in \ttbar\
production, and is interfaced to the {\sc Herwig} (v6.5) generator \cite{Herwig,Frixione:2010ra} for
simulation of the hadronization and fragmentation processes and to {\sc Jimmy} \cite{jimmy} for simulation of the underlying event from multiple parton interactions.
The {\sc Powheg} generator \cite{powheg} in combination with the {\sc Pythia} \cite{Pythia} or {\sc Herwig} generators is used for  
studying parton-shower systematic uncertainties.  For the study of other systematic uncertainties 
(top quark mass dependence, initial and final state radiation (ISR/FSR)), \ttbar\ samples produced with the {\sc Acermc} generator \cite{ACERMC} interfaced with
{\sc Pythia} 
are used.  
The expected \ttbar\ event yield is normalized to  the  cross-section of 164.6 pb, obtained with approximate next-to-next-to-leading-order (NNLO) QCD calculations \cite{hathor_10}.
Electroweak single-top-quark production is simulated using the MC@NLO generator and the event samples are normalized to approximate NNLO cross sections: 65 pb ($t$-channel) \cite{kidonakis_D83}, 
4.6 pb ($s$-channel) \cite{kidonakis_D81} and 15.7 pb ($Wt$ channel) \cite{kidonakis_D82}.

The background from \Wboson\ $+$ jets and \Zboson\ $+$ jets production is simulated with the {\sc Alpgen} v2.13 generator \cite{Alpgen}  and CTEQ6L1 \cite{Pumplin:2002vw} PDFs 
in exclusive bins of parton multiplicity for multiplicities of less than five,
and inclusively for five or more. The events are processed by {\sc Herwig} and {\sc Jimmy}.
The overall $\Wboson+$~jets and  $\Zboson+$~jets samples are normalized to the NNLO inclusive cross sections \cite{melnikov_06}.
Diboson samples are produced using {\sc Herwig} and {\sc Jimmy} with {\sc MRST2007LO} \cite{mrst2007lo} PDFs. 
 Dijet samples used for crosscheck purposes (see section \ref{sec:stat}) are generated using the {\sc Pythia} generator with the {\sc ATLAS AMBT2B} {\sc Pythia} tune \cite{PythiaAMBT2B} and with  {\sc MRST2007LO} PDFs.

\section{Event selection }
\label{sec:select}

The reconstructed events are selected using criteria designed to identify the lepton$\,+\,$jets final states, i.e. \ttbar\ events in which one of the \Wboson\ bosons decays leptonically and the other hadronically.
This sample also contains a significant fraction of \ttbar\ events where both \Wboson\ bosons decay leptonically, but one of the leptons is not reconstructed in the detector or fails the lepton identification requirements. 
In the simulated sample the events generated in both the single-lepton and dilepton channels are treated as signal if they satisfy the lepton$\,+\,$jets reconstruction criteria.

\subsection{Object reconstruction}
\label{objects}

An electron candidate is defined as an energy cluster deposition in the EM calorimeter associated with a well-reconstructed
charged particle track in the ID \citep{electronCP}. 
The candidate must have a shower shape consistent with expectations based on simulation, test-beam studies and $Z \ra ee$ events in data. The associated  ID track must satisfy quality criteria including the presence of high-threshold hits in the transition radiation tracker. All candidates are
required to have transverse energy ($\ET $) above 25 GeV and $|\eta | <$ 2.47, where $\eta $ is the pseudorapidity of the EM calorimeter cluster associated
with the electron. Candidates in the transition region between the barrel and end-cap calorimeters (1.37 $< |\eta | <$ 1.52) are excluded.

Muon candidates are reconstructed by combining track segments from different layers of the muon chambers \citep{ATLAS-CONF-2011-063}. 
Such segments are assembled starting from the outermost layer, with a procedure that takes material effects into account, and
are then matched with tracks found in the ID. The candidates are re-fitted exploiting the full track information from
both the muon spectrometer and the ID. They are required to have transverse momenta ($\pT $) above 20 GeV and the candidate muon must be within $|\eta | <$ 2.5.

Jet  candidates are reconstructed using the anti-$k_{t}$ algorithm~\citep{Cacciari_2008}
with jet radius parameter $R~=~0.4$. These jets are calibrated to the hadronic energy scale, using a $\pT $- and 
$|\eta |$-dependent correction factor obtained from simulation, test-beam and collision data ~\citep{jet_syst}.

The missing transverse momentum, $E\mathrm{_T^{miss}}$, is calculated as the magnitude of the vector sum of the energy deposits
in calorimeter cells associated with topological clusters \citep{topoCluster}, with the direction defined by the interaction vertex and position of the energy deposition in the calorimeter \citep{metCP}. 
The calorimeter cells are associated with a parent physics object in a chosen order:
electrons, jets and muons, such that a cell is uniquely associated with a single physics object. Cells belonging
to electrons are calibrated at the EM energy scale whereas cells belonging to jets are corrected to
the hadronic energy scale. Finally, the transverse momenta of muons passing the event selection are included, and the contributions from the calorimeter cells associated with the muons are subtracted.
The remaining clusters not associated with electrons or jets are included at the EM energy scale.

Overlap between the different object categories is avoided by the following procedure.
Jets within $\Delta R$ = 0.2 of an electron passing the electron selection requirements are removed from the list of jet candidates.\footnote{$\bigtriangleup R$ is defined  as a distance,
 $\bigtriangleup R = \sqrt{(\bigtriangleup \eta ^2 +\bigtriangleup \phi ^2)}$, in $\eta$-$\phi $ space, where $\eta $ is the pseudorapidity and $\phi $ is the azimuthal angle around the beam pipe.}
Muons within $\Delta R$ = 0.4 of any jet with $\pT\ >$ 20 GeV are rejected.
In addition, if a selected electron is separated by less than $\Delta R$ = 0.4 from any jet with $\pT\ >$ 20 GeV, the event is rejected (for event selection see section~4.2).

Tracks used for the $b$-jet charge calculation (see section~\ref{methods}) are required to contain at least six hits in the silicon microstrip detector and at least one pixel hit.
Only tracks with $\pT\ >$ 1 GeV and  $|\eta | <$ 2.5 are considered.
In addition, proximity to the $pp$ collision primary vertex\footnote{The primary vertex is chosen as the reconstructed vertex with the highest $\sum{p\mathrm{_T^2}}$ of associated tracks. At least five tracks with $\pT >$ 0.4 GeV are required.} expressed in terms of impact parameter in the transverse plane, $d_{0}$, and along the beam direction, $z_\mathrm{0}$, and good track fit quality are also required. The applied selection requirements on $d_{0}$ and $z_{0}$ are $|d_{0}| < 2~\text{mm}$ and  
$|z_{0} \cdot \sin(\theta)| < 10~\text{mm}$, and that on the quality of the track fit is $\chi^{2} / \text{ndf} < 2.5$.

For all reconstructed objects in the simulation, corrections are applied to compensate for the difference in reconstruction
 efficiencies and resolutions between data and simulation.

\subsection{Selection of {\textbf{\textit{\ttbar}}}  candidates}

The \ttbar\ candidates in the electron$\,+\,$jets or muon$\,+\,$jets final states are first selected with
a single-electron or single-muon trigger with transverse energy or momentum thresholds at 20~\GeV\ or 18~\GeV, respectively.
Events passing the trigger selection are required to contain exactly one reconstructed lepton, with $\ET >$ 25 GeV for an electron or $\pT >$ 20 GeV for a muon. 
At least four jets with transverse momenta $p_{\mathrm{T}}>$~25~\GeV\ and within the pseudorapidity range $|\eta |<$~2.5 are required. The missing transverse momentum,  $E_\mathrm{T}^\mathrm{miss}$, 
has to exceed 35~GeV for the events with electrons, and 20~GeV for the events with muons. 
In addition, a primary vertex containing at least five charged particles with  $p_\mathrm{T} >$ 0.4 \GeV\ is
required, and events containing jets with $\pT >$ 20 GeV in poorly instrumented detector regions are removed.

The transverse mass of the leptonically
decaying \Wboson\ boson in the event is reconstructed as $m_\mathrm{T}(W)= \sqrt{2p\mathrm{^{\ell}_Tp^{\nu}_T} (1-\rm{cos}(\phi^{\ell}-\phi^{\nu}))}$, where
the measured $E_\mathrm{T}^\mathrm{miss}$ magnitude and direction provide the transverse momentum, $p^{\nu}_\mathrm{T}$, and azimuthal angle, $\phi^\nu$, of
the neutrino, and the superscript $\ell$ stands for the $e$ or $\mu$. For events with electrons $m_\mathrm{T}(W)$ has to exceed 25~\GeV, while the
sum of $m_\mathrm{T}(W)$  and $E_\mathrm{T}^\mathrm{miss}$ has to exceed 60~\GeV\ for the events with muons. 

Finally, at least one jet is required to be $b$-tagged using the $b$-tagging procedure described in Ref.
\citep{JetFitterCombNN}. The procedure combines an algorithm based on jet track impact parameters with respect to the primary vertex with an algorithm exploiting the topology of $b$- and $c$-hadron weak decays inside the jet. The combination of the two algorithms is based on artificial neural network techniques with MC-simulated training samples and variables describing the topology of the decay chain used as the neural network input \citep{JetFitterTag}.
The chosen $b$-tagging  operating point corresponds to a 70\% tagging efficiency for $b$-jets in simulated
 \ttbar\  events, while light-flavour jets are suppressed by approximately a factor of 100.

These selection requirements, common to most of the ATLAS \ttbar\ analyses (see e.g.~\citep{ttbarXs_2011}), are further referred to as the basic \ttbar\ requirements. They are followed by requirements specific for reconstruction of the $b$-quark charge. In order to use the track charge weighting method (see section~\ref{charge_weighting}), the presence of a second $b$-tagged jet is required.
Each of the two $b$-tagged jets has to contain at least two well-reconstructed tracks with transverse
momenta above 1~GeV within the pseudorapidity range $|\eta |<$~2.5. A pairing criterion between the lepton and a $b$-jet is also applied (see section~\ref{methods}).

\section{Top quark charge determination}
\label{methods}

\indent The correlation between the top or exotic quark charge and the charges of their decay products can be used for the quark charge determination.
In the SM the top quark is expected to decay according to
\begin{equation}
t^\mathrm{(2/3)}\rightarrow b^\mathrm{(-1/3)} + W^\mathrm{(+1)} ,
\label{r1}
\end{equation}
\noindent while the exotic quark ($t_\mathrm{X}$) with charge --4/3 is assumed to decay according to\\
\begin{equation}
t\mathrm{_{X}^{(-4/3)}}\rightarrow b^\mathrm{(-1/3)} + W^\mathrm{(-1)}, 
\label{r2}
\end{equation}
where the electric charges of the particles are indicated in parentheses.
\noindent Considering the subsequent leptonic decay of the \Wboson\ bosons, $W^{\pm }\rightarrow \ell^{\pm }+\nu_{\ell} (\bar \nu_{\ell})$, the expectation for the SM case is that a positively charged lepton $\ell^\mathrm{+}$ is associated with the $b$-quark ($Q_{b} = -1/3$) from the same top quark,
while for the exotic case it is just the opposite: $\ell^\mathrm{-}$ is paired with the $b$-quark.
In the SM case the product of charges of the top or anti-top quark decay products ($Q\mathrm{_{\ell^+}}\times Q_{b}$  or $Q\mathrm{_{\ell^-}}\times Q_{\bar{b}}$) always has a negative sign while in the exotic case the sign is positive.
 
The charge of the \Wboson\ boson is taken from the charge of the high-$p_\mathrm{T}$ lepton in the event.  The charge of the quark initiating the $b$-jet is estimated from a weighted average of the charges of the tracks in the jet (see section~\ref{charge_weighting}).  
A lepton--$b$-jet pairing criterion (hereafter referred to as $\ell b$-pairing) is then applied to match the \Wboson\ boson to the $b$-jet from the same top quark (see section~\ref{jet_pairing}).  

\subsection {Weighting procedure for {\textit{b}}-jet charge calculation}
\label{charge_weighting}

\noindent For the determination of the effective $b$-jet charge a weighting technique ~\citep{Field_Feynman, Aleph_b-charge} is applied in which
the $b$-jet charge is defined as a weighted sum of the $b$-jet track charges,

\begin{equation}
Q_{b\mathrm{-jet}}=\frac {\sum_{i} Q_{i} \vert \vec j \cdot \vec p_{i} \vert ^\mathrm{\kappa}}
{\sum_{i} \vert \vec j \cdot \vec p_{i} \vert ^{\mathrm\kappa}},
\label{r3}
\end{equation}

\noindent where $Q_{i}$ and $\vec{p}_{i}$ are the charge and momentum of the $i$-th track, $ \vec j$ defines the $b$-jet axis direction, and $\kappa$ is a  parameter which was set to be 0.5 for the best separation between $b$- and $\bar {b}$-jets mean charges using the standard MC@NLO \ttbar\ simulated sample.

The calculation of the $b$-jet charge uses a maximum number of ten tracks with $\pT\,>$ 1~GeV associated with the $b$-jet within a cone of $\Delta R\,<$ 0.25. 
The $b$-jet tracks used in the calculation of the effective $b$-jet charge include not only the charged decay products of the $b$-hadron, but also $b$-fragmentation tracks, and can possibly also contain tracks from multiple interactions or pile-up.
The mean number of charged tracks within the $b$-jet cone is six for \ttbar\ $b$-jets. If there are more than ten associated tracks, the highest-\pT\ tracks are chosen. The maximum number of tracks, the minimum track \pT\ and the value of  $\Delta R$ were optimized using the standard MC@NLO \ttbar\ simulated sample. 
 The optimization takes into account that the pile-up effect can be stronger for the high track multiplicity events and that low-\pT\ tracks, coming mainly from gluons, could dilute the jet charge.

The variable that is used to distinguish between the SM and exotic model scenarios is the combined lepton--$b$-jet charge (hereafter referred to as the combined charge) which is defined as
\begin{equation}
Q_{\mathrm{comb}}=Q_{b-\mathrm{jet}}^\mathrm{\ell}\cdot Q_{\mathrm{\ell}},
\label{Comb}
\end{equation}
where $Q_{b\mathrm{-jet}}^\mathrm{\ell}$ is the charge of the $b$-jet  calculated with equation~(\ref{r3}) \footnote{The superscript $\ell$ is added to $Q_{b\mathrm{-jet}}$ to stress that the $b$-jet is paired with a lepton.} and $Q_\mathrm{\ell}$  the charge of the lepton, the two being associated via the $\ell b$-pairing described below.

\subsection {Lepton and {\textit{b}}-jet pairing algorithm}
\label{jet_pairing}

The $\ell b$-pairing is based on the invariant mass distribution of the lepton and the $b$-jet, $m(\ell,\mbox{$b$-jet})$. If the assignment is correct, 
assuming an ideal invariant mass resolution, $m(\ell,b\mbox{-jet})$ should not exceed the top quark mass provided that the decaying particle is the SM top quark.
Otherwise, if the lepton and $b$-jet are not from the same decaying particle, there is no such restriction. This is shown in Figure~\ref{fig1}, where the invariant mass distribution of a lepton and a $b$-jet in the signal MC sample is plotted for the correct pairing and the wrong pairing, for events fulfilling the basic \ttbar\ selection requirements.
For MC events the reconstructed $b$-jet is paired with a parton-level $b$-quark if their separation $\Delta R$ is less than 0.2; similarly, $\Delta R <$ 0.2 is required for the matching between parton-level and reconstructed leptons.
\begin{figure}[!h]
\centering
\epsfig{file=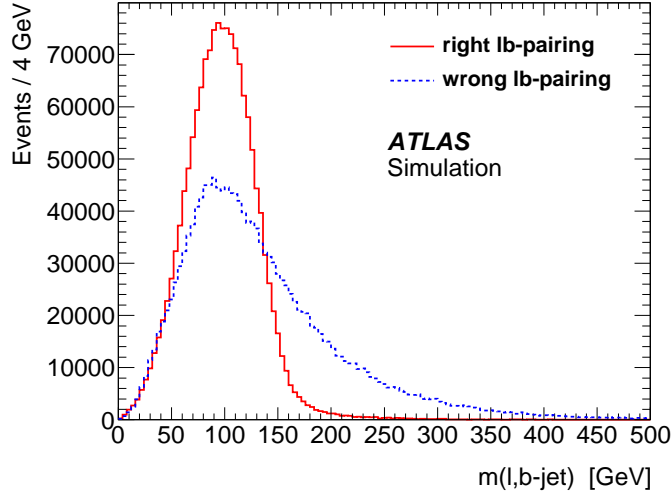, height=7cm}
\caption{Lepton--$b$-jet invariant mass spectra  for the lepton and $b$-jet pairs from the same top quark (right pairing, solid red line) and
for those originating from two different top quarks (wrong pairing, dashed blue line).}
\label{fig1}
\end{figure}

The $\ell b$-pairing requires events with two $b$-tags and only the events with $b$-jets that satisfy the conditions:

\begin{eqnarray}
  m(\ell,b\mbox{-jet}_{1})<m_\mathrm{cut} & \mbox{and} & m(\ell,b\mbox{-jet}_{2})>m_\mathrm{cut} \nonumber  \\
         &\quad \mbox{~or~} \quad & \\
  m(\ell,b\mbox{-jet}_{2})<m_\mathrm{cut} & \mbox{and} & m(\ell,b\mbox{-jet}_{1})>m_\mathrm{cut} \nonumber 
\label{r5}
\end{eqnarray}

\noindent are accepted. Here $b$-jet$_1$ and $b$-jet$_2$ denote the two $b$-tagged jets ordered  in descending order of transverse momentum. The optimal value for the $\ell b$-pairing mass cut, $m_\mathrm{cut}$,
is a trade-off between the efficiency ($\epsilon$) and purity ($P$) (see section~\ref{mc_reconstruction}) of the $\ell b$-pairing method. 
It was found by maximizing the quantity $\epsilon (2P-1)^2$ which is largest and nearly constant in the region 140~\GeV\ to 165~\GeV.
The value for the $\ell b$-pairing mass cut is chosen to be $m\mathrm{_{cut}}=155$~\GeV. A similar interval for the optimal value of $m_\mathrm{cut}$ was obtained using the relative uncertainty of the mean combined charge as an alternative figure of merit in the optimization. 

The efficiency of the $\ell b$-pairing procedure, defined as the ratio of the number of $\ell b$-pairs after and before the invariant mass cuts in equation 5.5,  is small ($\epsilon$=28\%), but it gives a high purity ($P$=87\%). The efficiency of the full set of selections used in the analysis, with respect to the basic \ttbar\ requirements, is reduced not only by the $\ell b$-pairing conditions but also by the requirement of the second $b$-tag (70\% efficiency) and, to a lesser extent, by the $b$-jet track requirements (see section~\ref{objects}) with efficiency around 99\%.

\section{Signal and background expectations}
\label{mc_optimisation}


The sensitivity for determining the SM top quark charge in the lepton$+$jets channel is investigated using MC and data control samples with the aim of finding the $Q_{\mathrm{comb}}$ expectations for the SM signal and background. 
Both single-lepton ($\ttbar\rightarrow \ell\nu jj\bbbar $) and dilepton ($\ttbar\rightarrow \ell\nu \ell\nu b\bar{b} $) 
samples are included for the signal. 

\subsection{Reconstructed signal distribution}
\label{mc_reconstruction}

In the MC analysis of the top quark charge the MC@NLO, {\sc Powheg} and {\sc Acermc} \ttbar\ samples are used. MC@NLO is taken as the default generator.  
The $b$-jet charge spectra reconstructed for the \ttbar\ electron$\,+\,$jets events from MC@NLO are presented in figure~\ref{fig:b-ch_MC-IM}. The distributions of $Q_\mathrm{{b-jet}}$  for $b$-jets paired with positive and negative leptons are shown after the $\ell b$-pairing.
In addition, the $Q_\mathrm{{comb}}$ spectrum (see equation~(\ref{Comb})) is also shown in the plot. 

\begin{figure}[!h]
\centering
\begin{tabular}{cc}
\epsfig{file=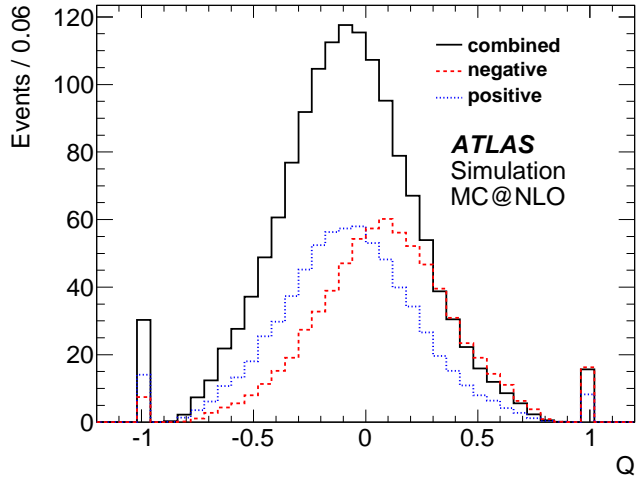, height=7cm, clip=}
\end{tabular}
\caption{Distributions of the reconstructed $b$-jet charge in electron$\,+\,$jets  \ttbar\ events (MC@NLO) associated with positive (dotted blue line) and negative (dashed red line) leptons and the combined charge (solid black line) after  the $\ell b$-pairing is applied. Here $Q$ represents  $Q_{b\mathrm{-jet}}^\mathrm{\ell}$ in the first two distributions and $Q_\mathrm{{comb}}$ in the third one.}
\label{fig:b-ch_MC-IM}
\end{figure}
The peaks at $\pm$1 in figure \ref{fig:b-ch_MC-IM} correspond to the cases where all the tracks within the $b$-jet cone of $\Delta R =$ 0.25 have charges of the same sign. In these cases the weighting procedure (equation~(\ref{r3}))  gives $Q_{b\mathrm{-jet}} = \pm$1.

The difference between the mean $b$-jet charges associated with $\ell^+$  and $\ell^-$ is clearly seen in figure~\ref{fig:b-ch_MC-IM}. 
 The results of the MC $b$-jet charge analysis are summarized in table~\ref{mc_bjetQ}, where the mean combined charges and charge purities are shown for different MC generators and the individual lepton$\,+\,$jets channels. The uncertainties in the mean combined charges 
of all MC samples are downgraded to the integrated luminosity of 2.05~\ifb\ corresponding to the size of the processed data sample.
The charge purity, $P_\mathrm{Q}$, is defined as
\begin{equation}
P_\mathrm{Q}=\frac{N(Q_{\mathrm{comb}}<0)}{N(Q_{\mathrm{comb}}<0)+N(Q_{\mathrm{comb}}\geq 0)},
\label{PQ}
\end{equation}
where $N(Q_{\mathrm{comb}}<0)$ and $N(Q_{\mathrm{comb}}\geq0)$ denote the number of events with $Q_{\mathrm{comb}} <0$ and $Q_{\mathrm{comb}}\geq0$, respectively. It is an important parameter which defines the quality of the $b$-jet charge weighting procedure. The higher $P_\mathrm{Q}$ is relative to 50\%, the better the flavour tagging identification is, i.e. the ability to distinguish between jets initiated by $b$- and $\bar{b}$-quarks. As shown in table \ref{mc_bjetQ}, our procedure produces $P_\mathrm{Q}$ near 60\%. 

In general, as it follows from table \ref{mc_bjetQ}, there is good agreement among the MC@NLO, {\sc Powheg} and {\sc Acermc} results on $Q_{\mathrm{comb}}$. The combined (electron $+$ muon channels) expectations agree to within 4\%. Good agreement is also seen between the individual channels.

\begin{table*}[htb]
 \begin{center}
 \begin{tabular}{c|c|c|c}
 \hline\hline
 Generator       		  & Channel    &   $\langle Q_\mathrm{comb}\rangle$     & $P_{Q}$ \\ \hline
						  & $e$        &    -0.0802 $\pm$ 0.0065   &   0.610 $\pm$ 0.003\\
   MC@NLO      			  & $\mu $     &    -0.0776 $\pm$ 0.0058   &   0.603 $\pm$ 0.003\\              
						  & $e+\mu$    &    -0.0787 $\pm$ 0.0043   &   0.606 $\pm$ 0.002\\ 
\hline                                                                                               
						  & $e$        &    -0.0739 $\pm$ 0.0070   &   0.595 $\pm$ 0.010\\
{\sc Powheg}+{\sc Herwig} & $\mu $     &    -0.0787 $\pm$ 0.0063   &   0.600 $\pm$ 0.008\\                   
						  & $e+\mu$    &    -0.0766 $\pm$ 0.0047   &   0.602 $\pm$ 0.006\\ 
\hline                                                                                               
						  & $e$        &    -0.0824 $\pm$ 0.0068   &   0.613 $\pm$ 0.010\\
{\sc Powheg}+{\sc Pythia} & $\mu $     &    -0.0703 $\pm$ 0.0063   &   0.594 $\pm$ 0.008\\ 
   						  & $e+\mu$    &    -0.0756 $\pm$ 0.0046   &   0.602 $\pm$ 0.006\\ 
\hline                                                                                               
						  & $e$        &    -0.0728 $\pm$ 0.0065   &   0.598 $\pm$ 0.011\\	
{\sc Acermc}+{\sc Pythia} & $\mu $     &    -0.0786 $\pm$ 0.0058   &   0.609 $\pm$ 0.008\\
   						  & $e+\mu$    &    -0.0760 $\pm$ 0.0043   &   0.604 $\pm$ 0.007\\  \hline\hline

\end{tabular}
 \bigskip
 \caption{The expected mean  combined charges ($\langle Q\mathrm{_{comb}}\rangle$) and charge purities ($P_\mathrm{Q}$) for the electron ($e$), muon ($\mu$) and combined ($e+\mu$) channels compared for the \ttbar\ MC@NLO, {\sc Powheg}+{\sc Herwig}, {\sc Powheg}+{\sc Pythia} and {\sc Acermc}+{\sc Pythia} simulated signal at 7 TeV in the lepton$\,+\,$jets channel obtained with the $\ell b$-pairing. The $\langle Q\mathrm{_{comb}}\rangle$ values are shown with their statistical uncertainties scaled to the integrated luminosity of 2.05~\ifb\ (see text).
The uncertainty of $P_\mathrm{Q}$ is obtained from the full MC sample and is not downgraded to the integrated luminosity of the data as $P_\mathrm{Q}$ reflects the quality of the charge weighting procedure.
}
\label{mc_bjetQ}
\end{center}
\end{table*}
To evaluate the effect of the reconstruction on the combined charge
, the mean associated $b$-jet charge reconstructed using the $\ell b$-pairing is compared with that based on the correct association of the lepton and $b$-jet using a MC generator-level matching. 
The comparison is carried out using the MC@NLO \ttbar\  samples
and the results  are shown in table~\ref{mc_invMass} for the  electron$\,+\,$jets, muon$\,+\,$jets and combined electron$+$muon channels. 
\begin{table*}[htb]
 \begin{center}
 \begin{tabular}{l|ccc}
 \hline\hline
 {Pairing type}   &         $e$                & $\mu $                         & $e + \mu $              \\ \hline
 MC matching          & -0.1014 $\pm$ 0.0009       &  -0.1006 $\pm$ 0.0008          &  -0.1010 $\pm$ 0.0006  \\  			 
 $\ell b$-pairing     & -0.0802 $\pm$ 0.0008       &  -0.0776 $\pm$ 0.0007          &  -0.0787 $\pm$ 0.0005  \\
 \hline\hline
\end{tabular}
 \bigskip
\caption{Comparison of  the mean combined charge, $\langle Q\mathrm{_{comb}}\rangle$, for the electron ($e$), muon ($\mu $) and combined ($e+\mu $) channels obtained using the MC matching and $\ell b$-pairing. The charges are shown with their statistical uncertainties for the full \ttbar\ MC@NLO  sample.}
\label{mc_invMass}
\end{center}
\end{table*}
The larger value of the average $Q\mathrm{_{comb}}$ for the MC matching  can be explained by its 100\% pairing purity. Table~\ref{mc_invMass} shows that
the expected mean combined charges obtained for the electron and muon channels are compatible within statistical errors for the MC matching. In the $\ell b$-pairing case a difference of 2.4$\sigma$ between the electron and the muon channel is seen. The difference can be explained by the non-identical selection criteria used for these two channels  and by the slight dependence of the $\ell b$-pairing efficiency and purity on lepton and $b$-jet transverse momentum. 
To illustrate that the analyzed sample of data does not have sufficient statistical power to be sensitive to such a difference, 
the statistical uncertainty quoted in table \ref{mc_bjetQ} has been scaled to the
luminosity of the analyzed data sample (2.05~\ifb).

\subsection{Background}
\label{bkgd}

The main background processes for the top quark charge measurement in the lepton$\,+\,$jets channel are: \Wboson $+\,$jets
production (the most significant background), 
 \Zboson $+\,$jets, multi-jet, diboson and single-top-quark production.
The single-top-quark background gives the same sign of the mean $b$-jet charge as the signal.
The MC simulation is expected to predict correctly all the processes with the exception of the multi-jet production and the normalization of the \Wboson $+\,$jets production. Though the probability for a multi-jet event to pass the event selection is very low,
the production cross section is several orders of magnitude larger than that of top quark pair production, and due to fake leptons\footnote{Fake lepton refers to both a non-prompt lepton and  a jet misidentified  as a lepton.} the multi-jet events can contribute to the background.
This background is determined in a data-driven way employing the so-called  Matrix Method \cite{ttbarXs_2011}. 
 This technique is based on the determination of the number of data events passing the full set of analysis selection criteria (tight selection) and that for a looser selection obtained by dropping the isolation requirement on the lepton. Using the number of events passing the tight and loose selections and the efficiencies for true and fake leptons, the number of fake-lepton events passing the tight \ttbar\ selection criteria is found. The efficiencies  are determined using appropriate control samples as is explained in detail in ref.~\cite{ttbarXs_2011}.  
The estimation of the \Wboson $+\,$jets background relies to a large extent on MC simulation, which is assumed to correctly describe the kinematics of the individual \Wboson $+\,$jets channels, but the overall normalization and flavour fractions are determined from data. 
The \Wboson $+\,$jets background is divided into four flavour groups: \Wboson $+b\bar{b}+$jets, \Wboson $+c\bar{c}+$jets, \Wboson $+c+$jets and \Wboson $+$light-flavour-jets. 
The flavour composition of the jets is determined from data based on the fraction of \Wboson  $+\,$jet(s) events that have one or two tagged jets \cite{Aad2012418}. 
The MC predictions for the \Wboson $+b\bar{b}+$jets and \Wboson $+c\bar{c}+$jets components are scaled by a factor of 1.63$~\pm~$0.76, the \Wboson $+c+$jets component by a factor of 1.11$~\pm~$0.35, and the light-flavour  \Wboson $+$jets component by a factor of 0.83$~\pm~$0.18 (for details see ref.~\cite{ttbarDifXs_2013}) .

The expected results for the electron and muon channels  after all selections used in the analysis,
including those used for the $\ell b$ pairing, are shown in table~\ref{tab:ExpBkg_emu}.
\begin{table*}[htb]
 \begin{center}
{\small
 \begin{tabular}{c|c|c|c|c}
\hline\hline
\multicolumn{1}{c|}{ }& \multicolumn{2}{c}{Electron } & \multicolumn{2}{c}{ Muon}  \\ 
\multicolumn{1}{c|}{{Process}}& \multicolumn{1}{c}{$N_{\mathrm\ell b}$}& \multicolumn{1}{c}{$\langle Q_\mathrm{comb}\rangle$} & \multicolumn{1}{c}{$N_{\mathrm\ell b}$} & \multicolumn{1}{c}{$\langle Q_\mathrm{comb}\rangle$}     \\ \hline
\Wboson\ $+$ jets    	&  77 $\pm$ 15  &  -0.077 $\pm$ 0.050	&  132 $\pm$	 23  & -0.047 $\pm$ 0.032  \\
\Zboson\ $+$ jets		&   9 $\pm$	 3  &   0.078 $\pm$ 0.153 	&   15 $\pm$	  4  & -0.179 $\pm$ 0.086 \\
Diboson					&   1 $\pm$  1  &  -0.229 $\pm$	0.573	&    2 $\pm$   	  2  & -0.071 $\pm$ 0.279	 \\
Multi-jet (DD)			&  18 $\pm$	18  &  -0.018 $\pm$	0.082	&   36 $\pm$	 36  & -0.027 $\pm$ 0.028	 \\
\hline
Non-top-quark background& 105 $\pm$	24  &  -0.015 $\pm$	0.041 	&  185 $\pm$	 43  & -0.052 $\pm$ 0.028	 \\\hline
Single-top-quark		&  67 $\pm$ 11  &  -0.066 $\pm$ 0.042	&   80 $\pm$  	 12  & -0.051 $\pm$	0.038   \\\hline
Signal					& 1420$\pm$ 150 &  -0.080 $\pm$ 0.007	& 1830 $\pm$ 	190  & -0.078 $\pm$ 0.006	\\ \hline
Signal + background		& 1600$\pm$ 150 &  -0.075 $\pm$	0.006	& 2100 $\pm$ 	200  & -0.074 $\pm$ 0.006	 \\
\hline \hline
\end{tabular}
}
 \bigskip
 \caption{Signal and background expectation after applying the $\ell b$-pairing separately for the electron and muon channels for 2.05 \ifb\ integrated luminosity. Here, DD stands for ``data driven", $N_{\mathrm\ell b}$  is the mean number of lepton--$b$-jet pairs and $\langle Q_\mathrm{comb}\rangle$ is the reconstructed mean combined charge. The non-top-quark background is the total background not including single-top-quark events. The uncertainties include the statistical uncertainties and the uncertainties in the cross sections and integrated luminosity.  }
\label{tab:ExpBkg_emu}
\end{center}
\end{table*}
The uncertainties in the expected number of the signal and background events include not only the statistical uncertainties but also the cross-section uncertainties, which vary from 10\% for signal and single-top-quark production to 100\% for the multi-jet background, and the uncertainty in the integrated luminosity (1.8\%).



\section{Results}
\label{result}

The distributions of the reconstructed quantities involved in the top quark charge determination,
namely the distributions of $b$-jet and lepton $p\mathrm{_T}$, $E\mathrm{_T^{miss}}$ and the number of tracks with $\pT>$~1~\GeV\ in a $b$-jet, were compared to the expectations after applying the basic \ttbar\ selection requirements and after  the full set of the analysis requirements including two $b$-tags and $\ell b$-jet pairing.
Fairly good agreement between data and MC distributions is observed. An example is seen in figure~\ref{fig:pTcomp}, which shows the $b$-jet \pT\ distribution after the basic \ttbar\ requirements and after the full set of the analysis requirements.
 \begin{figure*}[htb]
	\centering
	\begin{tabular}{cc}
		\epsfig{file=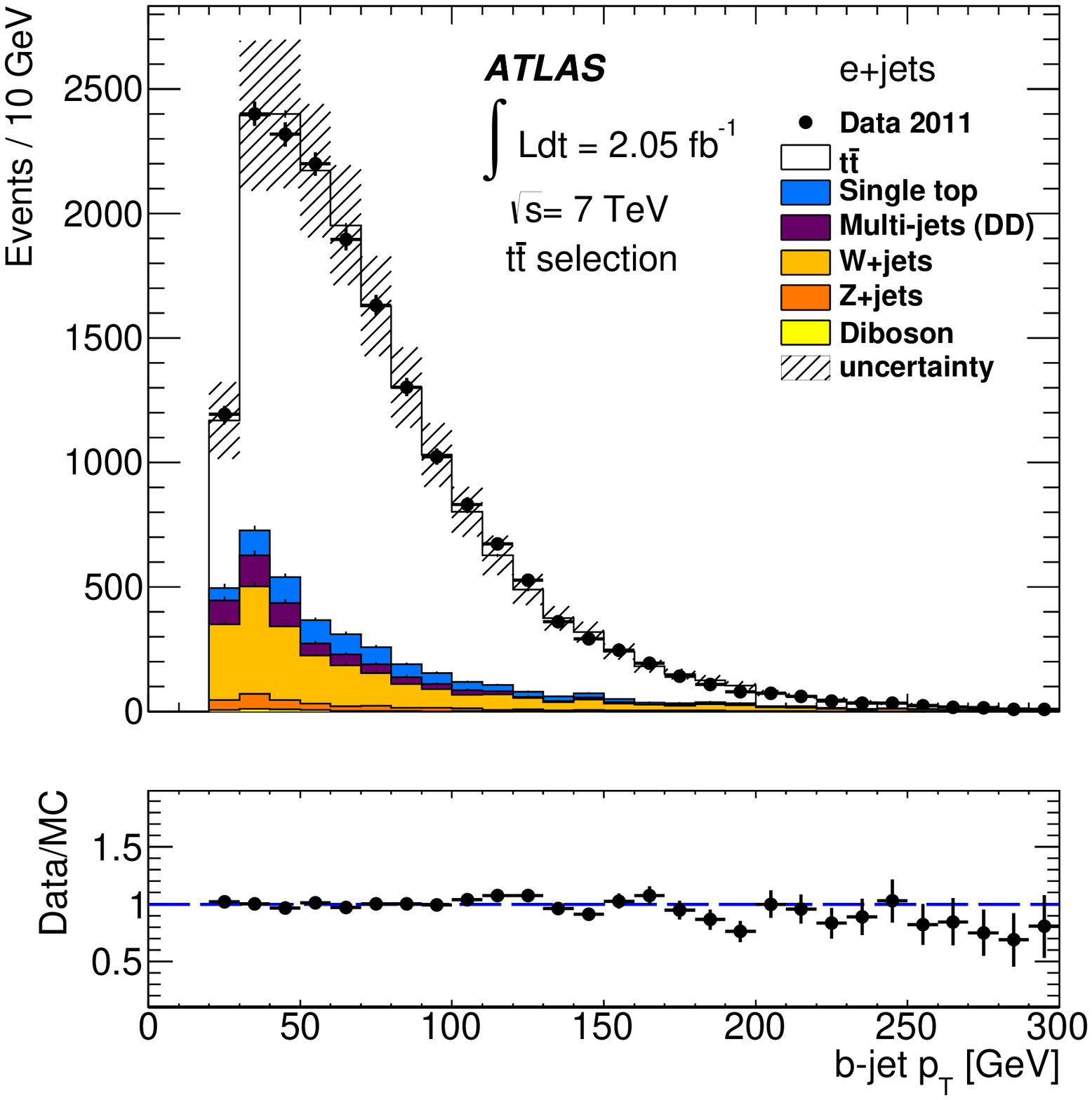,height=7cm, clip=}
		\epsfig{file=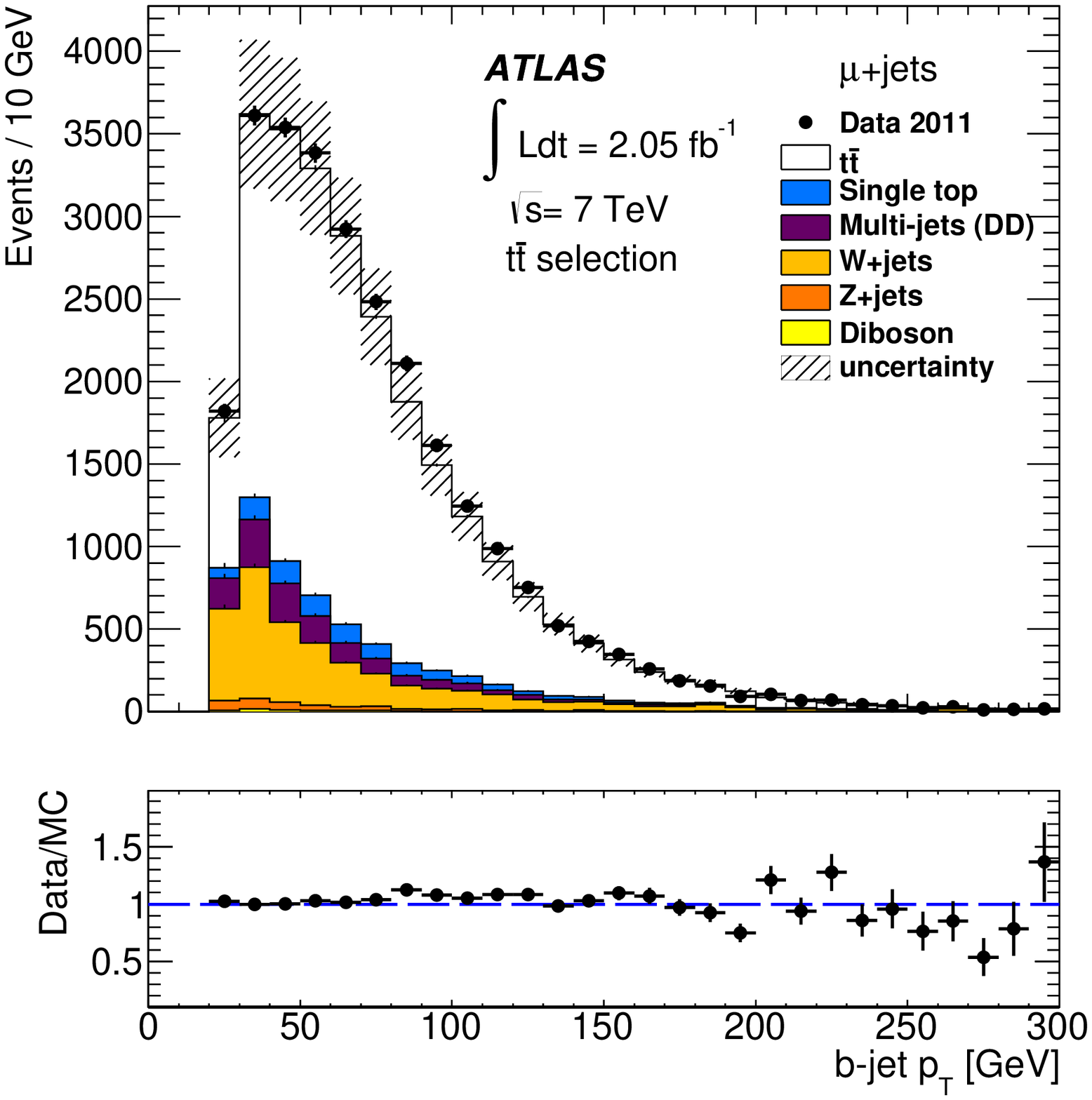, height=7cm,   clip=}\\
        \epsfig{file=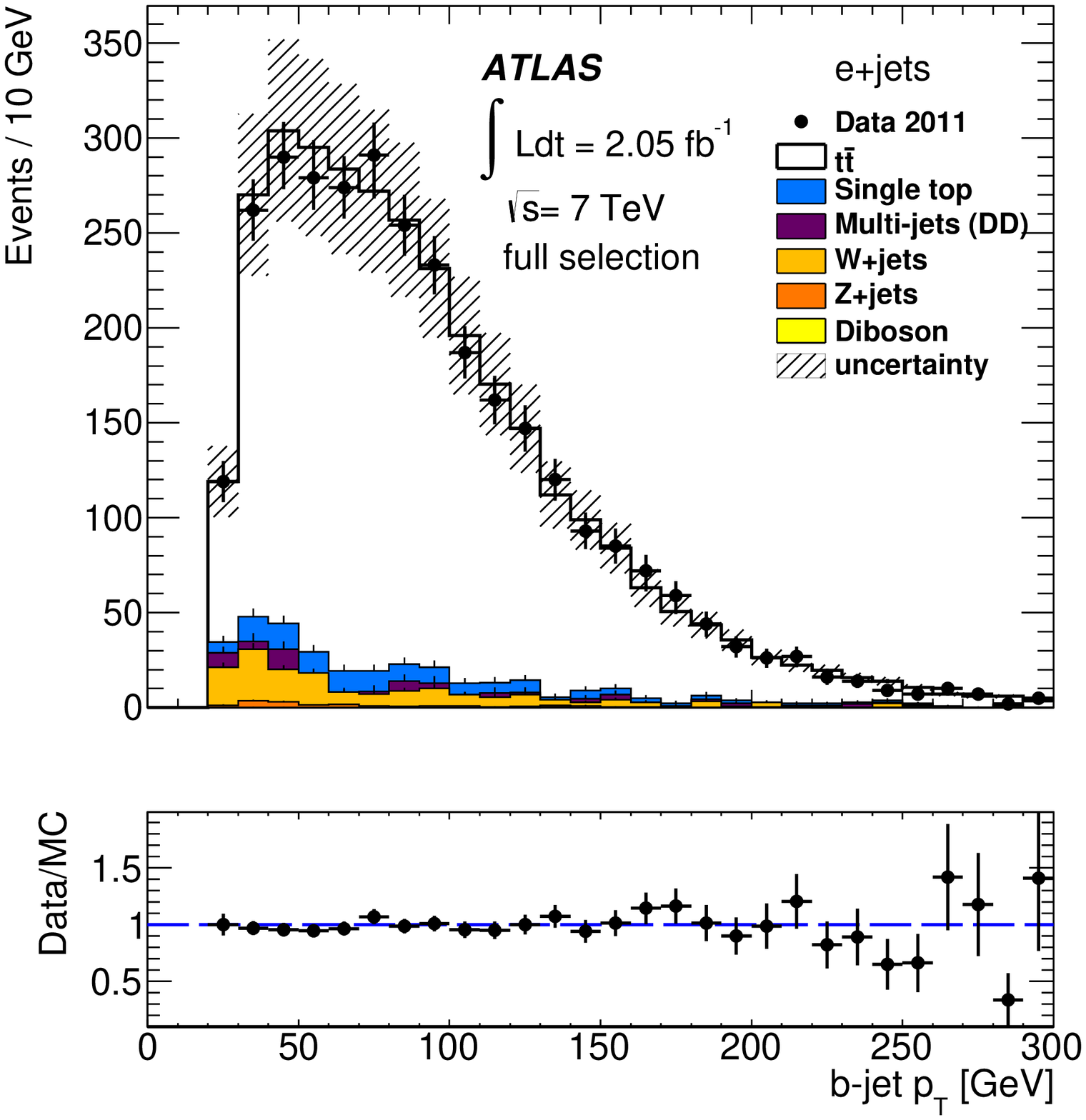, height=7cm, clip=} 
		\epsfig{file=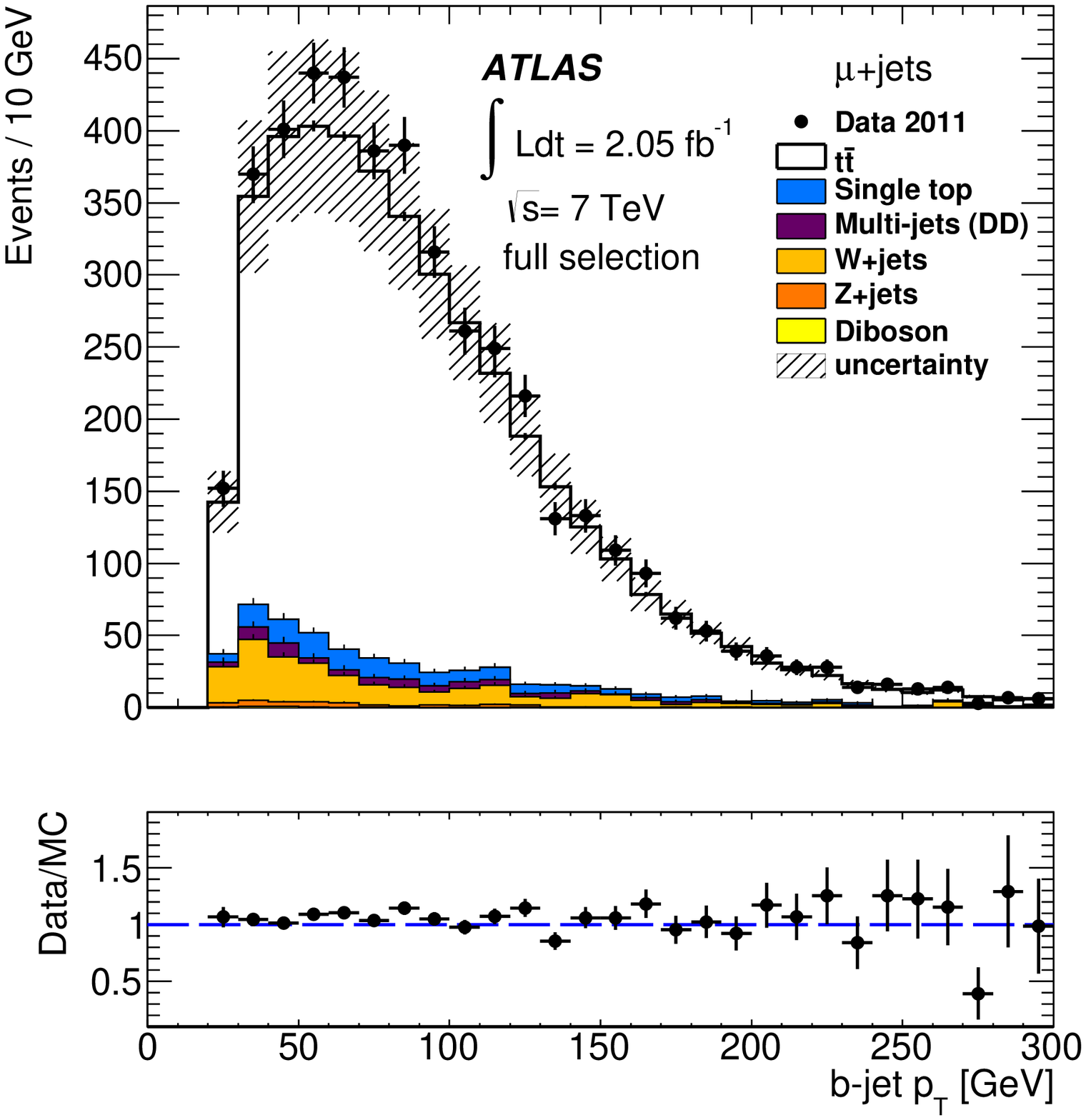, height=7cm, clip=}\\
	\end{tabular}
	\caption{Data and MC comparison of the  $b$-jet $p_\mathrm{T}$ distribution after the basic \ttbar\ requirements (upper plots) and after the full set of requirements (bottom plots) for electron$\,+\,$jets (left) and muon$\,+\,$jets (right) events. The MC expectations for signal and background are normalized to 2.05~\ifb\ using the expected cross sections. The shaded area belongs to the MC distribution and corresponds to a combination of the statistical uncertainties and the uncertainties in the cross sections and the integrated luminosity.}
	\label{fig:pTcomp}
\end{figure*}

To test the $b$-jet charge weighting procedure (see eq.~\ref{charge_weighting}), the reconstructed distributions of  the mean value of the absolute $b$-jet charge, shown as a function of $b$-jet $p_\mathrm{T}$  for the \ttbar\ candidate events in data and MC simulation, are compared in figure~\ref{fig:bQ_pT} after the basic \ttbar\ requirements and after  the $\ell b$-pairing. The expected background is subtracted from the data distribution. 
The distributions in figure~\ref{fig:bQ_pT} are profile histograms containing in each bin the mean value with its uncertainty depicted as the corresponding error bar.  Due to the high statistics of the MC samples, the error bars of the MC distributions are within the symbol size.
 Good agreement between the data and the MC simulation is observed. 
 An advantage of using the absolute value of $b$-jet charge is that it can be used for comparison of data and MC in different stages of the candidate event selection while the combined charge is available only after the full set of selection criteria. The relation between the mean combined charge and the mean value of absolute $b$-jet charge was investigated in a dedicated MC study, which showed a linear dependence.
In addition, figure~\ref{fig:bQ_pT} demonstrates that
the mean $b$-jet charge depends only weakly on the $b$-jet $p_\mathrm{T}$, especially for the distributions after the $\ell b$-pairing,  
which makes the charge weighting procedure insensitive to uncertainties in the $b$-jet $\pT$ distribution.
\begin{figure*}[htb]
	\centering
	\begin{tabular}{cc}
		\epsfig{file=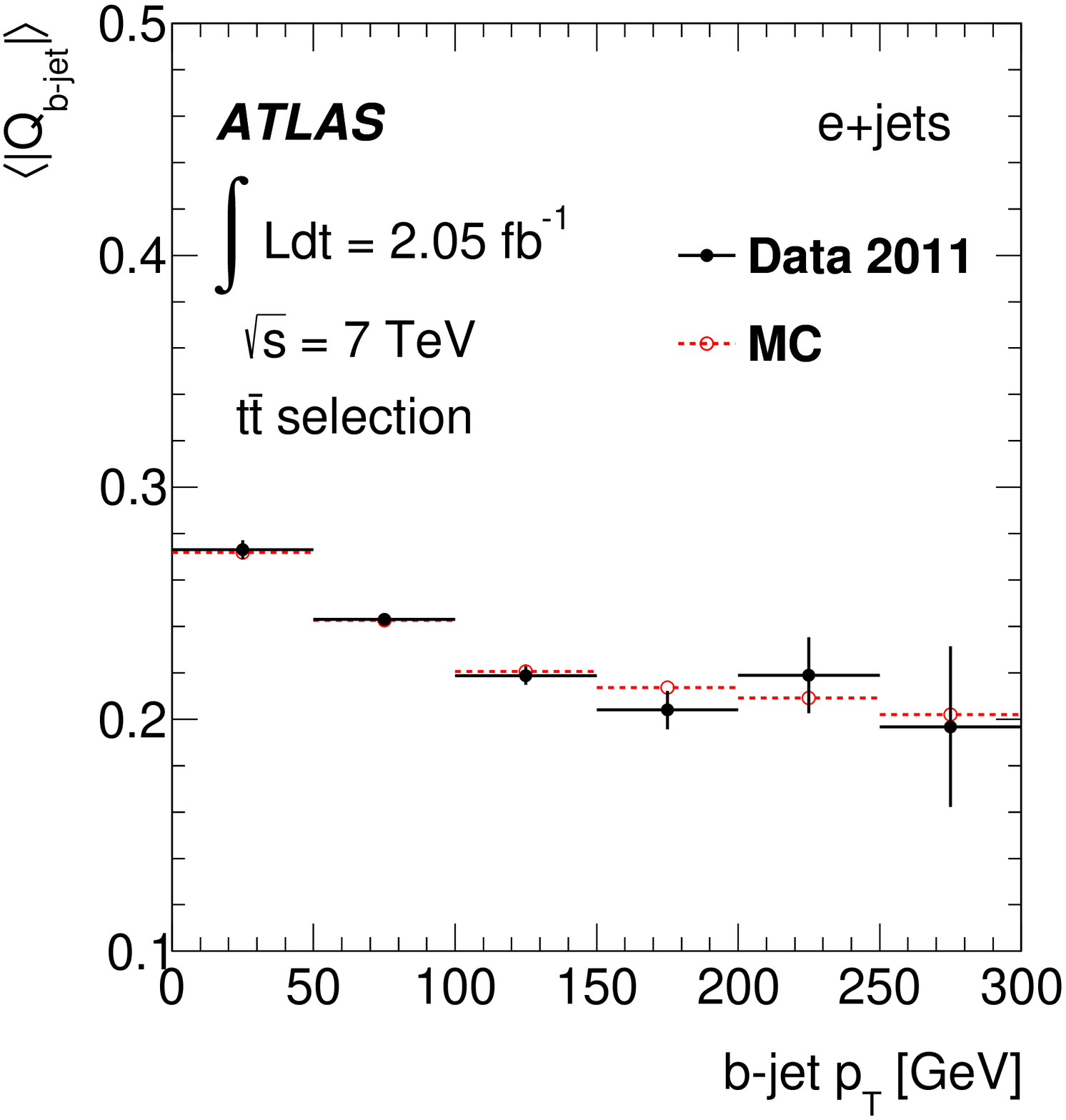,height=7cm, clip=}
		\epsfig{file=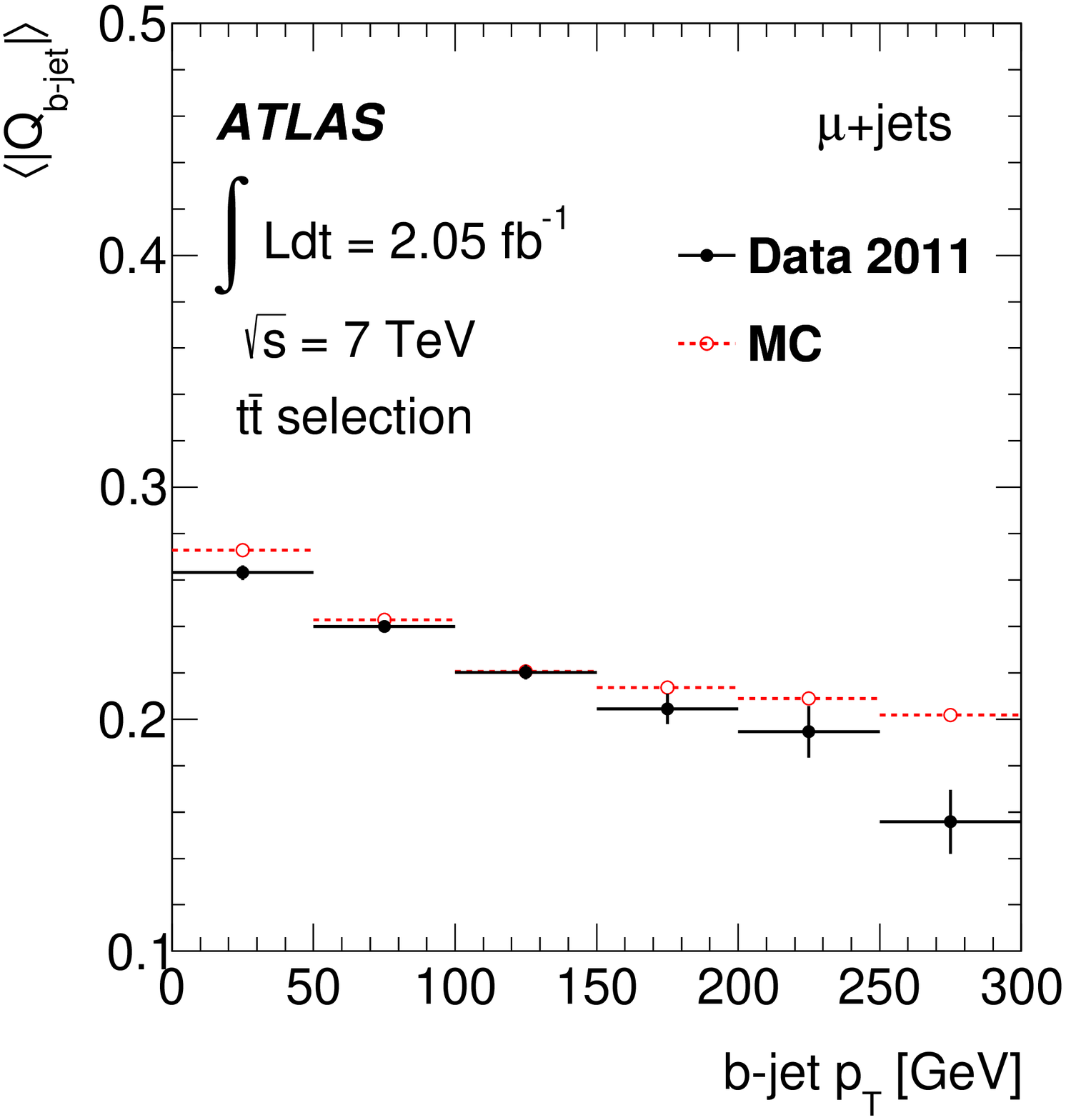, height=7cm,   clip=}\\
		\epsfig{file=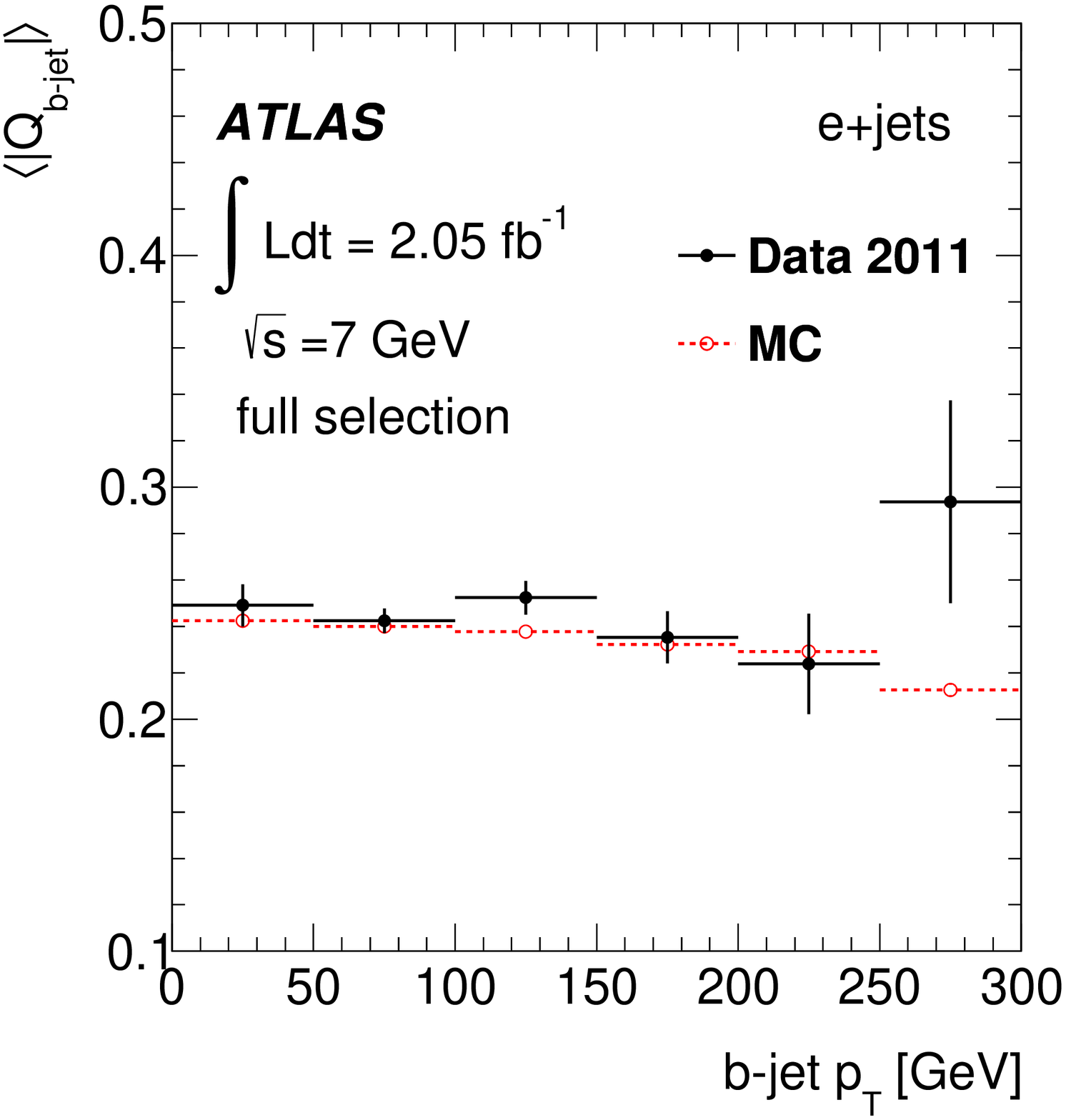, height=7cm, clip=}
		\epsfig{file=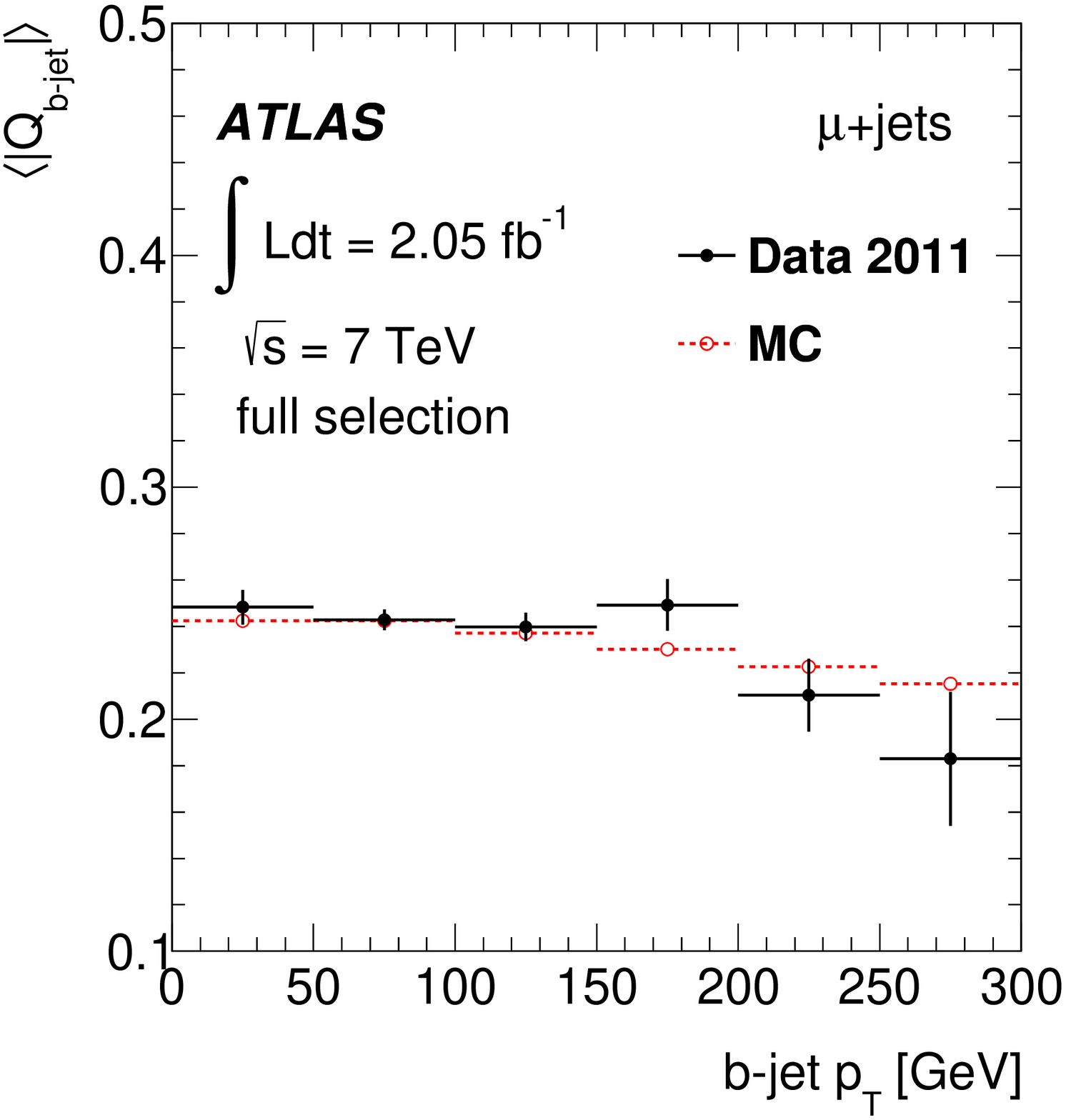, height=7cm, clip=} \
	\end{tabular}
	\caption{Data and MC comparison of the mean of the absolute value of the $b$-jet charge, $\langle\mid {Q}_{b\mathrm{-jet}}\mid \rangle$, as a function of $b$-jet $p_\mathrm{T}$ after the basic \ttbar\ requirements (upper plots) and after the full set of requirements (bottom plots) for electron$\,+\,$jets (left) and muon$\,+\,$jets (right) events. The data are shown after subtraction of the expected background and MC stands for MC@NLO \ttbar\ events. Only statistical uncertainties are shown.}
	\label{fig:bQ_pT}
\end{figure*}

The increasing instantaneous LHC luminosity was accompanied by an increasing mean number of reconstructed $pp$ interaction vertices per bunch crossing.
This quantity, which is a measure of pile-up (presence of additional interactions in the event), increased from 6 to 17 during the analysed 2011 data-taking period.  To assess the impact of pile-up, 
the mean of the absolute value of $b$-jet charge, $\langle\mid {Q}_{b\mathrm{-jet}}\mid \rangle$,  
is reconstructed as a function of the number of reconstructed $pp$ interaction vertices for both the data and MC samples and with the full set of the $t\bar{t}$ requirements used in this analysis  including two $b$-tags and $\ell b$-pairing.
No dependence is observed for the level of pile-up present in the data sample, as shown by figure~\ref{fig:bjetQ_vtx} for the absolute value of $b$-jet charge. The same level of stability is observed for the combined charge as a function of the primary vertex multiplicity.
\begin{figure*}[htb]
	\centering
	\begin{tabular}{cc}
		\epsfig{file=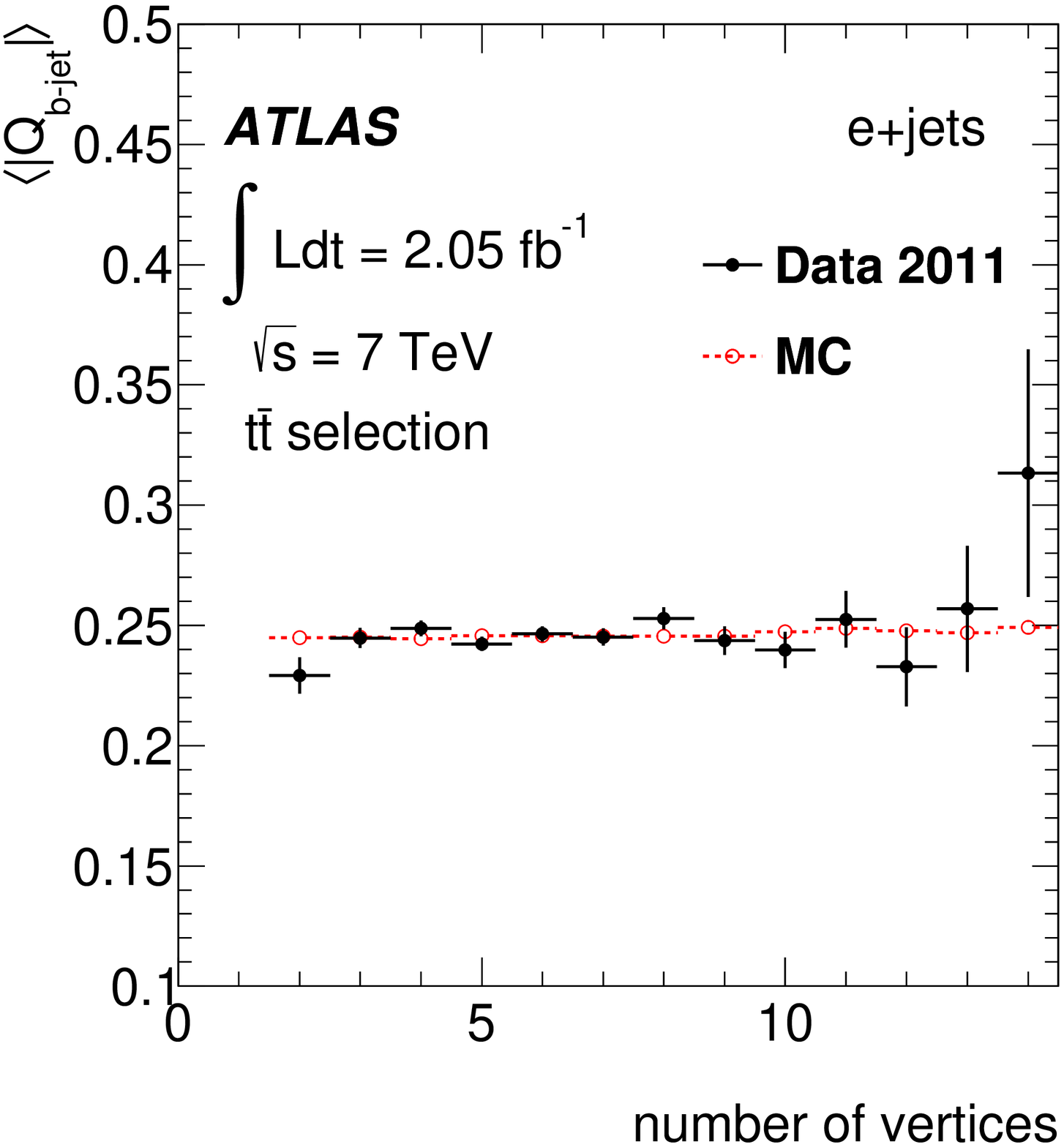, height=7cm, clip=}
		\epsfig{file=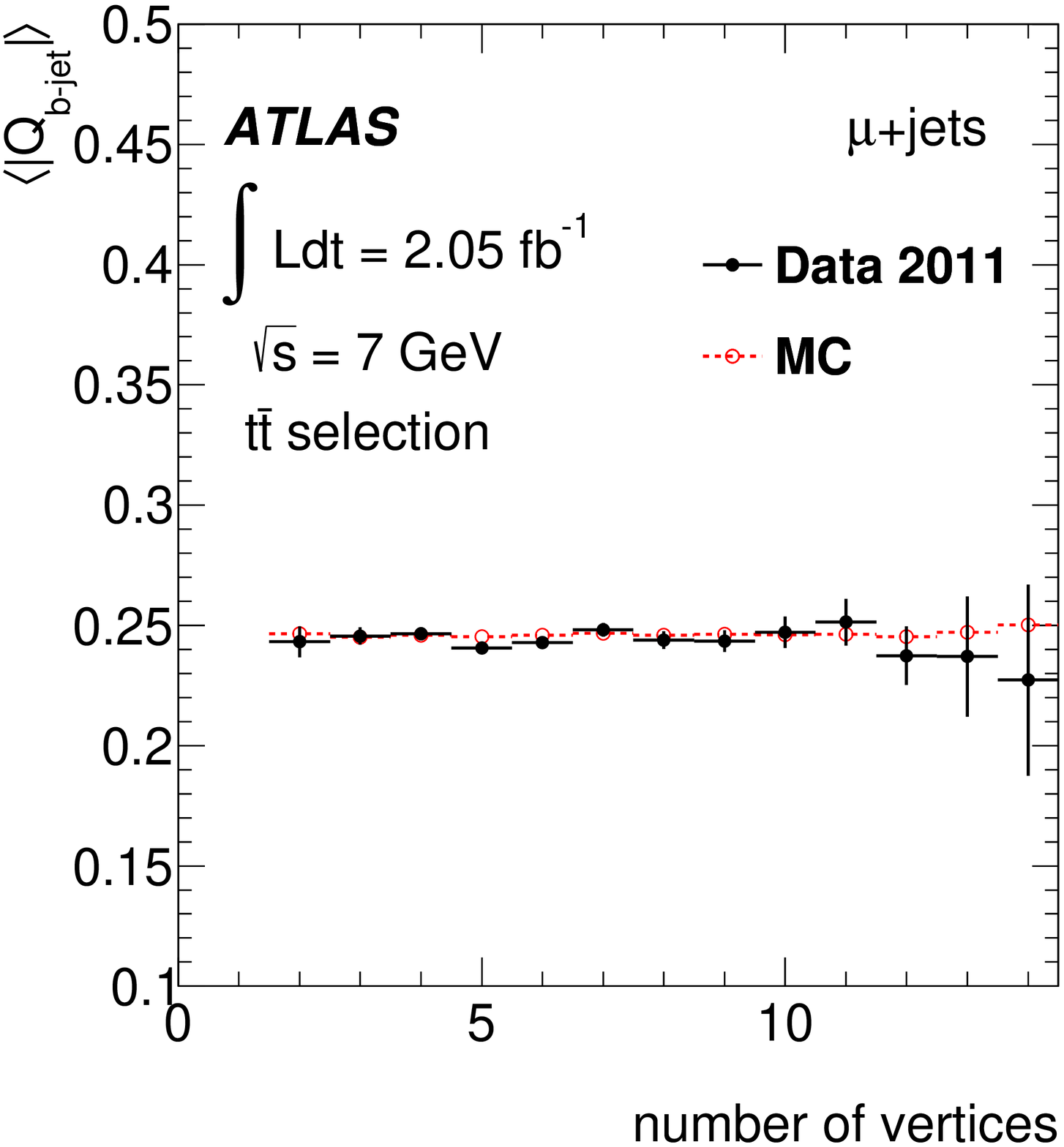, height=7cm, clip=}\\
	\end{tabular}
\caption{Data and MC (MC@NLO) comparison of the mean of the absolute value of the $b$-jet charge, $\langle\mid {Q}_{b\mathrm{-jet}}\mid \rangle$, as a function of vertex multiplicity after all the \ttbar\ requirements for electron$\,+\,$jets (left) and muon$\,+\,$jets (right) events.}
	\label{fig:bjetQ_vtx}
\end{figure*}

\begin{table*}[htb]
	\begin{center}
		\begin{tabular}{c|c|c|c|c|c}
			\hline \hline
\multicolumn{1}{c|}{Lepton } &$N_{\ell b}^{\mathrm{expect}}$& $N_{\ell b}^{\mathrm{data}}$ &\multicolumn{3}{|c}{ $\langle Q\mathrm{_{comb}}\rangle$ }   \\ \cline{4-6}
			channel  &    &    &  \multicolumn{1}{|c|}{SM expected} & \multicolumn{1}{|c|}{XM expected} &\multicolumn{1}{|c}{Data} \\ \hline
			$e$      &     1600 $\pm$ 150 & 1638 &	-0.075	$\pm$	0.006 & 	0.073	$\pm$	0.006 & -0.079	$\pm$ 0.008    \\
			$\mu$    &     2100 $\pm$ 200 & 2276 &	-0.074	$\pm$	0.006 & 	0.065	$\pm$	0.006 &	-0.075	$\pm$ 0.007    \\ 
			$e+\mu$  &     3700 $\pm$ 250 & 3914 &	-0.075	$\pm$	0.004 & 	0.069	$\pm$	0.004 & -0.077	$\pm$ 0.005    \\ \hline \hline

		\end{tabular}
		\caption{Number of $\ell b$-pairs expected from MC simulation ($N_{\ell b}^{\mathrm{expect}}$) and observed in data ($N_{\ell b}^{\mathrm{data}}$), and reconstructed mean combined charge, $\langle Q\mathrm{_{comb}}\rangle$, for the data in the different lepton$\,+\,$jets channels compared to those expected in the SM and the exotic model (XM). The uncertainties include the statistical uncertainties scaled to 2.05~\ifb\ and the uncertainties in the cross sections and integrated luminosity.}
		\label{data_Qcomb}
	\end{center}
\end{table*}

Figure~\ref{fig:Q_bjet} compares the $b$-jet charge spectra 
after the basic \ttbar\ cuts for the data and the expected sum of signal and background normalized to the integrated luminosity of 2.05~\ifb.  
The charge spectra are symmetric around zero and show good agreement between data and MC.
\begin{figure*}[htb]
	\centering
	\begin{tabular}{cc}
		\epsfig{file=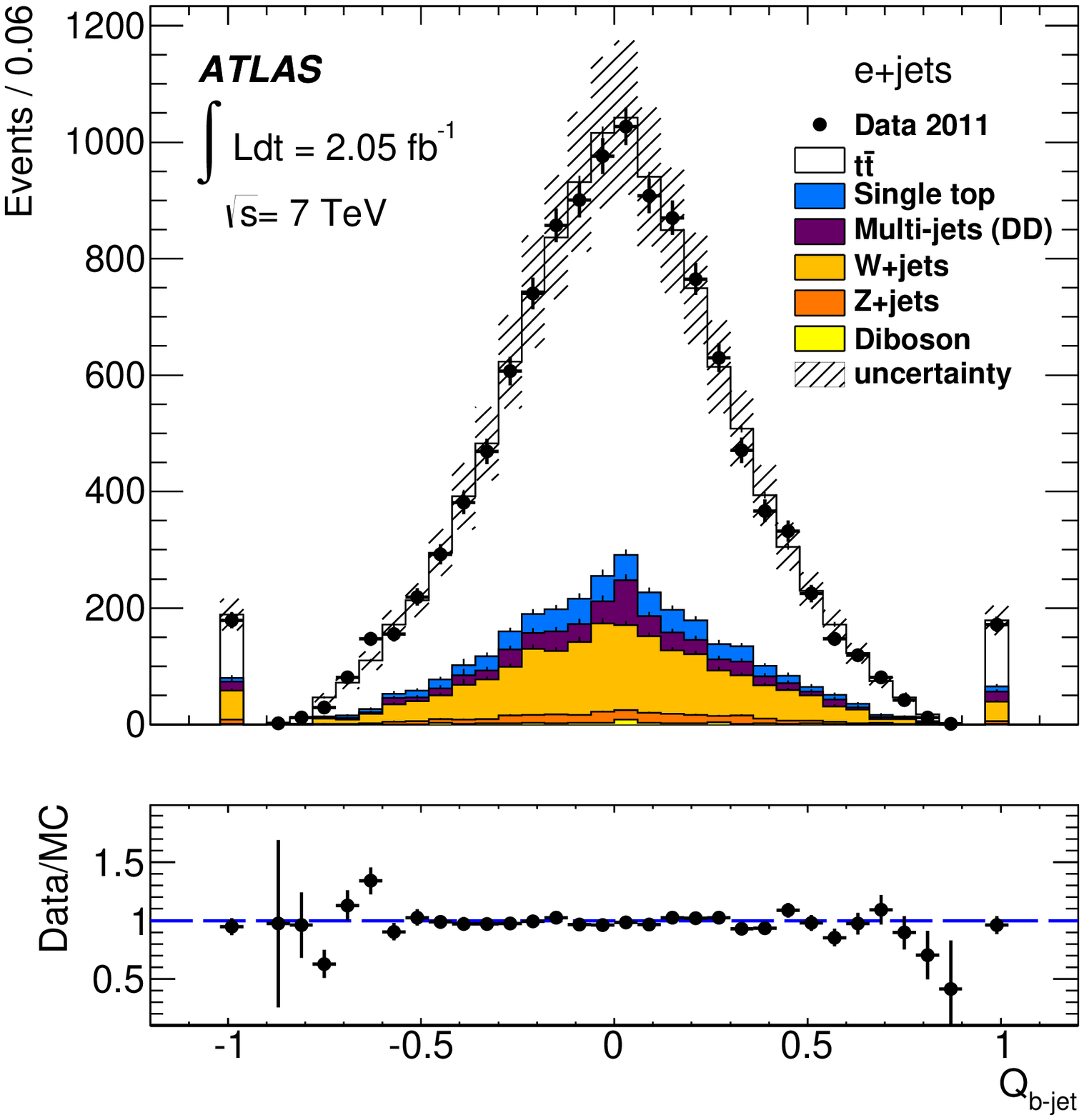,height=7.5cm, clip=}
		\epsfig{file=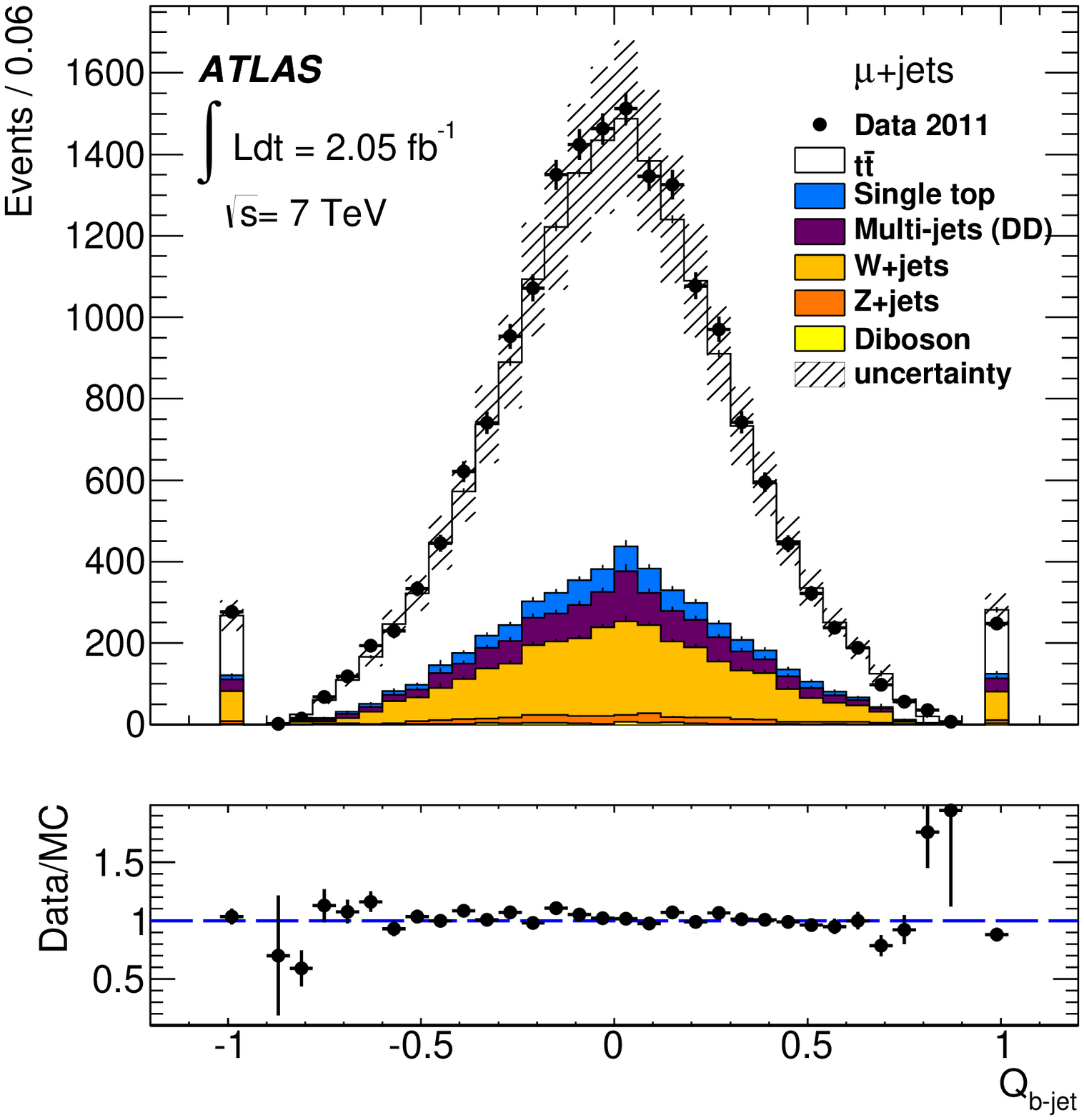, height=7.5cm,   clip=}\\
	\end{tabular}
	\caption{Data and MC comparison of the $b$-jet charge after the basic \ttbar\ requirements for electron$\,+\,$jets (left) and muon$\,+\,$jets (right) events. The MC expectations for signal and background are normalized to 2.05~\ifb\ using the expected cross sections. The shaded area corresponds to a combination of statistical uncertainties and uncertainties in the cross sections and integrated luminosity.}
	\label{fig:Q_bjet}
\end{figure*}

The results for the combined charge are summarized in table \ref{data_Qcomb}. This table contains the number of reconstructed lepton--$b$-jet pairs along with the  mean combined charge for the different channels.
The uncertainties  in the expected number of events in table \ref{data_Qcomb} include the cross-section uncertainty and the 1.8\% uncertainty in the integrated luminosity.

The combined charge for the exotic model in table~\ref{data_Qcomb} was obtained by inverting the signal \ttbar\  and single-top-quark combined charges  while the non-top-quark background charge was not changed. The inversion of the $b$-jet charge (or lepton charge) in a lepton--$b$-jet pair, provided that the lepton and $b$-jet come from a top quark decay, corresponds to a change of the decaying quark charge from 2/3 to --4/3. Such an approximation of the process with the exotic quark should be appropriate since the exotic quark differs from the top quark only in the electric charge. Although this could result in higher photon radiation in the exotic quark case, and consequently in a slightly softer $b$-jet \pT\ spectrum, this should not influence the combined charge since the photon radiation in the top quark case is only a small effect and the $b$-jet charge depends only weakly on $b$-jet \pT. This was verified by studying the exotic quark combined charge directly using events generated by {\sc Acermc}. The {\sc Acermc} sample gives, within statistical uncertainties, a compatible result with that obtained using the inversion procedure applied to the SM MC@NLO sample. 

From table \ref{data_Qcomb} it can be concluded that the data agree with the SM top quark hypothesis within the uncertainties and that the observed and expected numbers of events are also consistent with each other.
Figure \ref{fig:Qcomb_data-vs-MC} compares  the reconstructed combined charge spectra for the data with MC expectations for signal and background after $\ell b$-pairing  for  the electron$\,+\,$jets (left) and muon$\,+\,$jets (right) final states, showing good agreement between the data and the SM expectations.  
\begin{figure*}[htb]
	\centering
	\begin{tabular}{cc}
		\epsfig{file=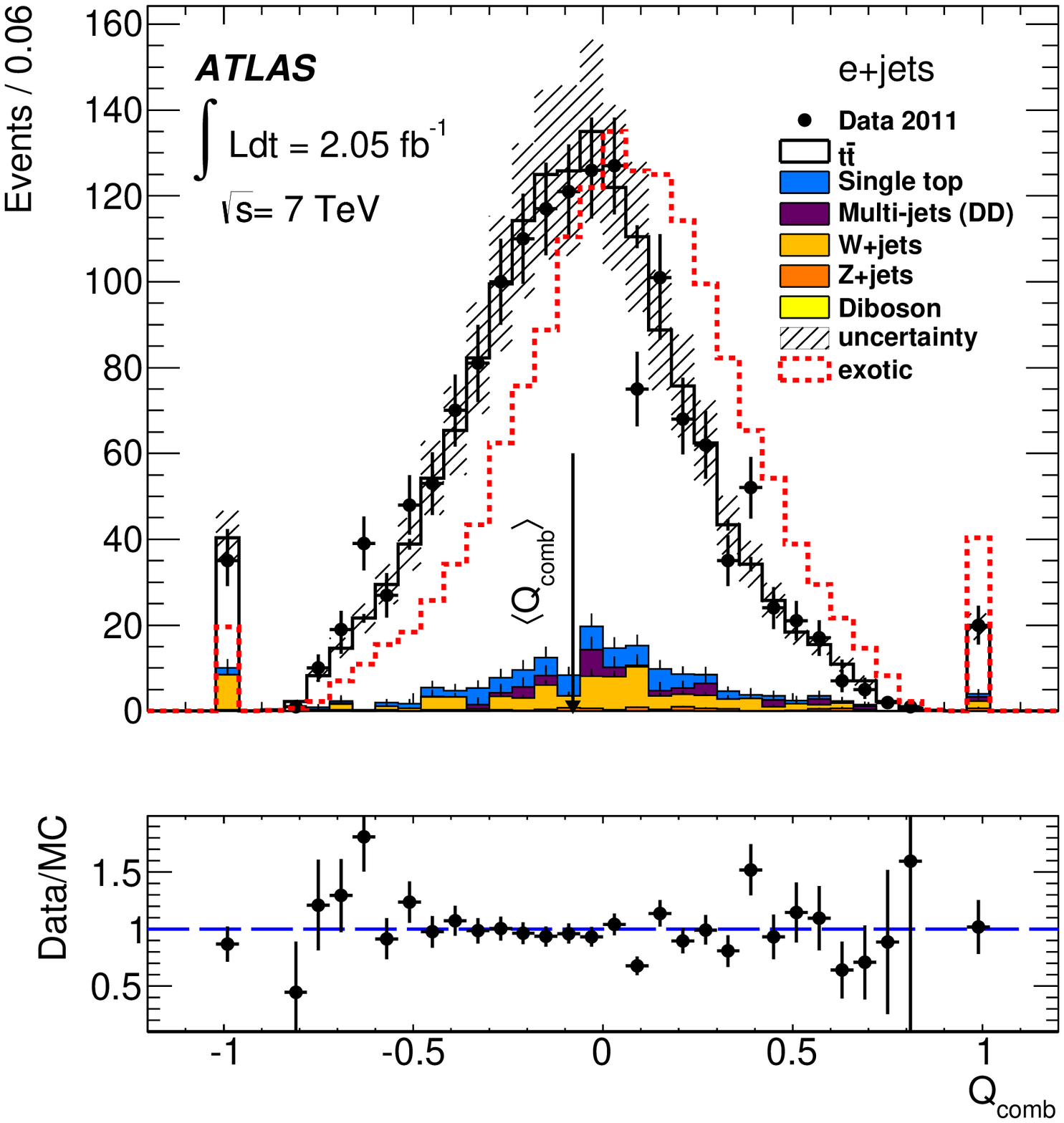, height=7.5cm, clip=}&
		\epsfig{file=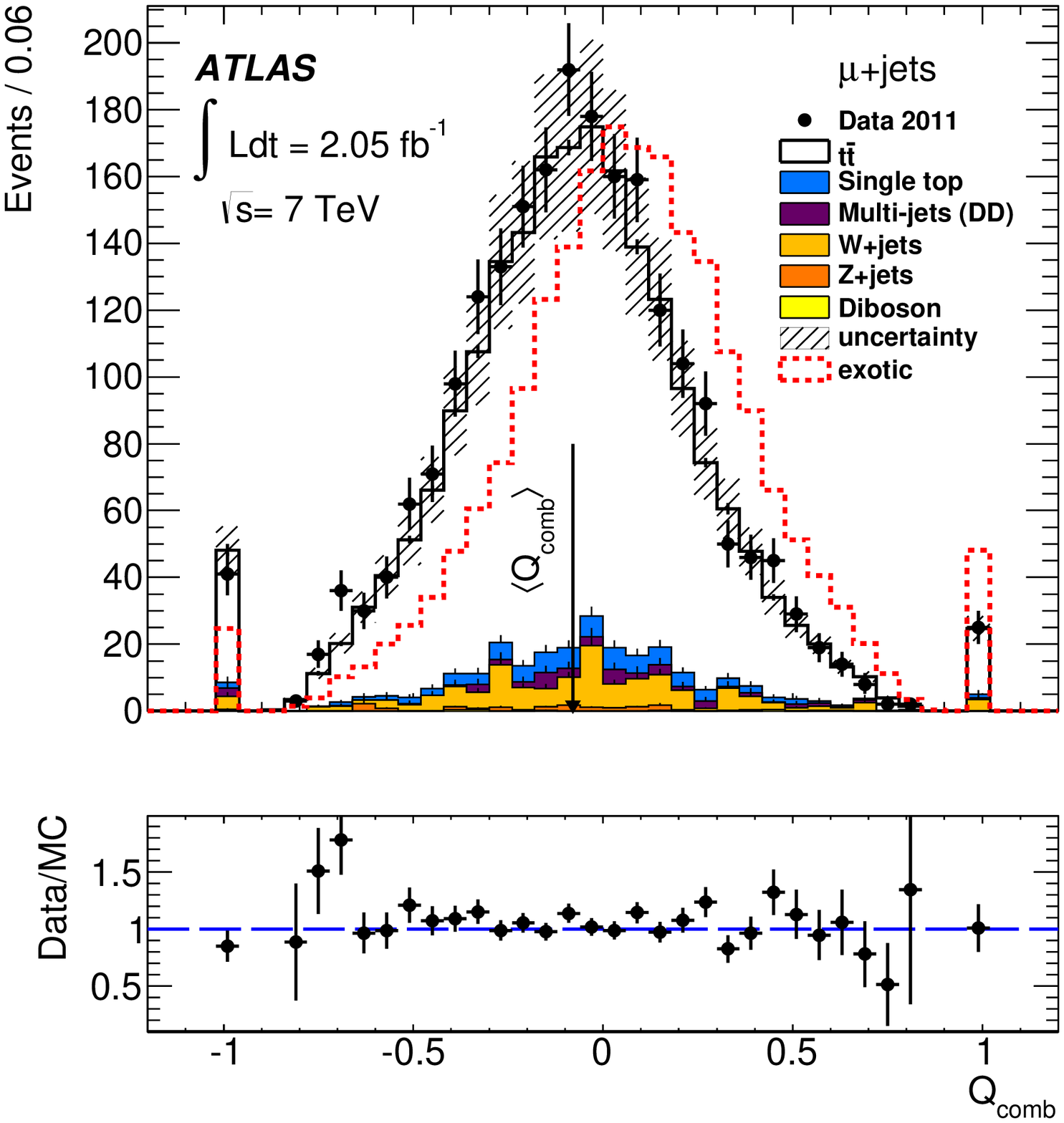, height=7.5cm,  clip=} \\
	\end{tabular}
	\caption{Distribution of the combined charge, $Q_\mathrm{comb}$, in electron$\,+\,$jets (left)
	and muon$\,+\,$jets (right) final states. The full circles with error bars are data, the full black line corresponds to the SM scenario, and the dashed red line corresponds to the exotic model. The vertical line, labeled with $\langle Q_\mathrm{{comb}}\rangle$, shows the mean value of the $Q_\mathrm{{comb}}$ distribution obtained from data. Only statistical uncertainties are shown.}
	\label{fig:Qcomb_data-vs-MC}
\end{figure*}

The top quark charge can be directly inferred from the background-subtracted \mbox{$Q_\mathrm{comb}$} data distribution 
using a $Q_\mathrm{{comb}}$ to $b$-jet charge calibration coefficient obtained from MC. 
From the SM value of the $b$-quark  charge ($Q_b =\mathrm{-1/3}$) and
the mean reconstructed value of the combined charge ($\langle Q_\mathrm{{comb}}\rangle$) for signal events, the $b$-jet charge calibration coefficient $C_\mathrm{b} = Q_\mathrm{b}/\langle Q_\mathrm{comb}\rangle$ is found to be $\mathrm{4.23\,\pm \,0.03}$~(stat.)~$\pm$~0.07~(syst.) when evaluated using the full \ttbar\  MC sample. The systematic uncertainty on $C_\mathrm{b}$ is taken as half the difference between the values of the calibration coefficient for the electron and muon channels. 
As mentioned in section~\ref{mc_reconstruction}
 the small difference between the mean combined charges of the electron and muon channels arises  as a consequence of  different selection criteria used for these channels. The mean combined charge depends slightly on $b$-jet $\pT$ and the  $\ell b$-pairing purity and efficiency depend on lepton and  $b$-jet $\pT$. Though these dependences are weak they should be taken into account if the common calibration coefficient is used.  
The top quark charge then can be calculated as

\begin{equation}
	Q_\mathrm{top}=1+Q_\mathrm{comb}^\mathrm{ (data)}\times C_b\ ,
\end{equation}
where  $Q\mathrm{_{comb}^{ (data)}}$ is the reconstructed $b$-jet charge obtained from the data after the subtraction of the expected background.

\begin{sloppypar}
The mean value of the top quark charge for the electron$\,+\,$jets channel is 
\mbox{$Q_\mathrm{top}$ = 0.63~$\pm$~0.04~(stat.) $\pm$~0.11~(syst.)} and that for the muon$\,+\,$jets channel is \mbox{$Q_\mathrm{top}$ = 0.65~$\pm$~0.03~(stat.)~$\pm$~0.12~(syst.)}.
The combined result using both channels is 0.64~$\pm $~0.02~(stat.)~$\pm$~0.08~(syst.). This result is obtained from the mean of the combined histogram of $Q\mathrm{_{comb}}$ for the two channels. 
The quoted systematic uncertainty includes uncertainties on the calibration constant and all the uncertainties on the mean combined charge as described below. 
\end{sloppypar}

\subsection{Systematic uncertainties}
\label{syst_e}

The studies of systematic uncertainties connected with the combined charge follow methods similar to those used in other top quark studies (see e.g. ref.~\citep{ttbarXs_2011}).
Each systematic effect is investigated by varying the corresponding quantity by $\pm 1\sigma $ with respect to the nominal value. 
 If the direction of the variation is not defined
(as in the case of the estimate resulting from the difference of two models, e.g. {\sc Herwig} and {\sc Pythia}), the estimated variation is assumed to be the
same size in the upward and the downward direction and the uncertainty on $\langle Q_\mathrm{comb}\rangle$ is symmetrized. The  following effects are taken into account.

{\it Monte Carlo generators} -- the systematic uncertainties from MC generators are
estimated by comparing the results obtained with the {\sc MC@NLO} and {\sc Powheg} generators.

{\it Showering and hadronization} -- the {\sc Powheg} samples with shower models from {\sc Pythia} or {\sc Herwig} are compared and the difference
      is taken as the uncertainty due to the showering model.
      
{\it Top quark mass} -- the uncertainty resulting from the assumed top quark mass 
      is estimated using simulated \ttbar\ samples with top quark mass in the range of 167.5--177.5 \GeV\
      in steps of 2.5 \GeV. After fitting the mean values  of $Q_\mathrm{comb}$ for different top quark mass samples the quoted systematic uncertainty is
      the largest of the differences between the fit function value at 172.5 \GeV\ and at those at 172.5 $\pm$ 1.0 \GeV.
      
{\it Initial- and final-state radiation (ISR/FSR)} -- the ISR/FSR uncertainty is calculated using dedicated signal samples generated with  {\sc Acermc} interfaced to {\sc Pythia}. 
      The parameters responsible for the level of ISR and FSR are varied in a range comparable to those used in the Perugia MC tunes \citep{MC_Perugia}.
      Half of the difference between the minimum and maximum values of $\langle Q_\mathrm{comb}\rangle$ is taken as the systematic uncertainty due to ISR/FSR.

{\it Colour reconnection} -- 	the systematic uncertainty due to colour reconnection is determined using {\sc Acermc} interfaced to {\sc Pythia}. Two different colour reconnection effects are simulated as described in refs. \citep{MC_Perugia, color_rec1} and for each effect the difference in the reconstructed combined charge between two levels of the colour reconnection is found. The larger difference is taken as the systematic uncertainty. 

{\it Missing transverse momentum} --  $E\mathrm{_T^{miss}}$  is used in the event selection and can influence the reconstructed $Q_\mathrm{comb}$. The impact of a possible mis-calibration is assessed by changing the measured $E\mathrm{_T^{miss}}$  within its uncertainty. The systematic uncertainty of $E\mathrm{_T^{miss}}$ includes the energy scale of clusters not associated with jets, electrons or muons and the accuracy of the pile-up simulations.
The effect of a hardware failure in a part of the liquid-argon calorimeter is also taken into account. This uncertainty is assessed by varying the jet thresholds used for removing events with jets in  the dead calorimeter region.

{\it Multi-jet normalization} -- a 100\% uncertainty on the number
      of multi-jet events due to the data-driven method is assumed in calculating the uncertainty of $\langle Q_\mathrm{comb}\rangle$ connected with this normalization.
      
{\it Single-top-quark normalization} -- the cross sections of individual single-top-quark channels are simultaneously varied within their theoretical uncertainty by $\pm 1\sigma $
      and the largest difference in the combined signal and background $\langle Q_\mathrm{comb}\rangle$ with respect to the nominal one is quoted as the systematic uncertainty due to the single-top-quark production cross section \citep{single_top_syst}.
      
{\it \Wboson + jets} -- the \Wboson $+\,$jets cross section is varied within its theoretical uncertainty (the 
uncertainty for inclusive \Wboson\ production of 4\% and the additional uncertainty per each additional jet, of 24\%, are added in quadrature). 
The uncertainties on the shapes of \Wboson $+\,$jets kinematic distributions are assessed by varying several parameters, such as the minimum transverse momentum of the partons and the functional form of the factorization scale in {\sc Alpgen}. 
The \Wboson $+\,$jets samples are reweighted according to each of these parameters and the quadratic sum of the uncertainties for the individual parameters  is taken as the systematic uncertainty. Uncertainties connected with the scaling factors correcting the fractions of heavy flavour contributions in simulated \Wboson $+\,$jets samples are also taken into account. 

{\it \Zboson $+\,$jets} -- the same prescription as for the normalization of  \Wboson $+\,$jets events  is also applied to \Zboson $+\,$jets events.

{\it b-tagging} -- 
the $b$-tagging efficiency and mistag probabilities in data and MC simulation are not identical. To reconcile the difference, $b$-tagging scale factors
together with their uncertainties are derived per jet \citep{JetFitterTag, JetFitterCombNN}. They depend on the jet $p_\mathrm{T}$ and $\eta $ and the underlying
quark flavour. For the nominal result, the central values of the scale factors are applied, and the systematic uncertainty
is estimated by changing their values within their uncertainties.

{\it  Lepton-related uncertainty} --  this item comprises the uncertainties due to MC modelling of the lepton identification, trigger efficiency, energy scale and energy resolution.
Each simulated event is weighted with an appropriate scale factor (ratio of the measured efficiency to the simulated one) in order to reproduce the efficiencies seen in data. The uncertainties on the scale factors are included in the uncertainties on the acceptance values. Details can be found in ref.~\citep{ttbarXs_2011}.

{\it  Jet energy scale} --  the jet energy scale (JES) and its uncertainty are derived by combining information from test-beam
data, LHC collision data and MC simulations \citep{jet_scale_EPJC_11, jet_scale_PRD_12}. 
The dependence of the JES uncertainty on the $p_\mathrm{T}$  and $\eta$ of the reconstructed jet is used  to scale the energy of each jet up or down by one standard deviation in the used MC sample. 
These variations are also propagated to the missing transverse energy. An uncertainty contribution to the JES due to pile-up events is also taken into account. An additional uncertainty is applied exclusively to $b$-jets. For each $b$-jet matched to a parton level $b$-quark a \pT -dependent uncertainty ranging from 2.5\% for low-\pT\ jets to  0.76\% for high-$p_\mathrm{T}$ jets is used. 

The JES is the most significant source of systematic uncertainty. The reason is that changes in the JES have a large impact on the number of events with low-\pT\ $b$-jets  and the purity of the $\ell b$-pairing degrades at low $b$-jet \pT. The number of events at high and low JES varies with respect to the nominal scale by 25\% and 14\%, respectively.  

{\it  Jet energy resolution} --  the impact of the jet energy resolution  is assessed by smearing the jet
energy before performing the event selection. The energy of each reconstructed jet in the simulation is additionally smeared by a Gaussian function such that the width of the resulting Gaussian distribution includes the uncertainty on the jet energy resolution.

{\it  Jet reconstruction efficiency} --  the impact of the uncertainty in the jet reconstruction efficiency is evaluated by randomly
dropping jets from events and determining the variation of $\langle Q_\mathrm{comb}\rangle$ with respect to that of the nominal sample, following the prescription described in ref. \citep{jet_syst}. 

{\it  Influence of $b$-hadron fractions} -- in the hadronization process that leads to a $b$-jet, different $b$-hadrons can be formed and  the combined charge can depend on the $b$-hadron type. In addition, the mixing of $B^0$ and $B^0_{\rm S}$ mesons needs to be taken into account.  For the $b$-jets containing $B^0$-mesons, it leads to a smaller mean combined charge in comparison with the jets containing charged $B$ mesons. The effect for  jets containing $B^0_{\rm S}$ mesons, where the mixing probability is 50\%, 
should lead to zero  mean combined charge. The  measured mixing probabilities ($\chi _{\rm d}$ = 0.186 ($B^0$) and $\chi _{\rm S}$ = 0.5 ($B^0_{\rm S}$)) \citep{PDG2012} are used to find the effective values of the mean combined charge for $b$-jets with $B^0$ and $B^0_{\rm S}$ mesons.
A study based on MC simulation shows that the mean combined charge for $b$-jets with $b$-baryons is about 74\% of that for $b$-jets with $B^\pm $. 
The systematic uncertainty on the mean combined charge due to the uncertainties on the $b$-hadron production fractions, taken from Ref. [45], has been evaluated by varying independently  the production fractions for $B^0$ and $B^0_{\rm S}$ mesons and $b$-baryons by 1 standard deviation up and down and adding the individual contributions in quadrature.

 All other systematic uncertainties are small (less than 1\%). 
	A summary of all systematic uncertainties for the reconstruction of the combined charge in the electron and muon channels combined is shown in table \ref{jet_scale}.
	\begin{table*}[htb]
		\begin{center}
			\begin{tabular}{l|ccc}
				\hline
				Source 	  							         &   Systematic uncertainty (\%) \\ \hline
				MC statistics 								 & 0.7  		\\
				MC generator 								 & 3.7  	 	\\
				Parton shower 								 & 7.9  	 	\\
				Colour reconnection							 & 0.5  	 	\\
				ISR/FSR 									 & 3.1  	 	\\
				Top quark mass 								 & 0.3	 	\\
				Missing transverse energy                   & 0.8  	 	\\
				Jet energy scale 							& 8.3 		\\
				$b$-jet energy scale 						& 3.3 		\\
				Jet energy resolution 						& 1.0	    \\
				Jet reconstruction efficiency 				& 0.7 	 	\\
				$b$-tagging              					& 0.3 		\\
				Single top normalization					& $<$0.1	 	\\
				\Wboson\ $+$ jets  							& 1.2	 	\\
				\Zboson\ $+$ jets 							& 0.1	 	\\
				Multi-jet normalization 					& 1.0		\\
				Electron-related uncertainty				& 1.3 	 	\\
				Muon-related uncertainty	 				& 1.8 	 	\\
                $b$-hadron fractions						& 0.7		\\ \hline
				Total uncertainty of $e+\mu $-channel		& 13.2 	 	\\ \hline
			\end{tabular}
			\caption{The systematic uncertainties for the combined charge. The total uncertainty is calculated
			by adding the individual ones in quadrature.}

			\label{jet_scale}
		\end{center}
	\end{table*}

\section{Statistical comparison of the SM and exotic model}
\label{sec:stat}
The main result of this analysis -- the compatibility of the data with the SM hypothesis of the top quark charge of 2/3 -- was evaluated using a statistical model. 
This model is based on the Cousins--Highland approach \citep{CandH_1992}. 
The test statistic used for this purpose is the mean value of the combined charge. 
Due to finite detector resolution and finite sample size, the mean value of the combined charge observed in the experiment can be treated as one realization of 
a random variable, $\bar Q$, the distribution of which characterizes all possible outcomes of the experiment. This variable can be expressed as
\begin{equation}
\bar{Q}=(1-r_\mathrm{b}-r_\mathrm{t})\cdot Q_{\mathrm{s}}+r_{\mathrm{b}}\cdot Q_{\mathrm{b}}+r_{\mathrm{t}}\cdot Q_{\mathrm{t}},
\label{Qbar}
\end{equation}
where $Q_{\mathrm{s}}$, $Q_{\mathrm{b}}$ and $Q_{\mathrm{t}}$  are the combined charge mean values for the signal, background and single-top-quark processes, respectively,
and $r_\mathrm{b}$ ($r_\mathrm{t}$) is the fraction of the background (single-top-quark) events in the total sample of the candidate events.

The SM  acceptance (critical) region \citep{james_2006, cowan_1998} is defined as $\bar{Q} < 0$ ($\bar{Q} > 0$). The decision boundary $\bar{Q} = 0$ unambiguously determines the confidence level $\alpha $ (probability to exclude the SM scenario if it is true) and the so-called false negative rate $\beta $ (the probability of failing to reject the alternative hypothesis if it is true).
The quantities $Q_{\mathrm{s}}$, $Q_{\mathrm{b}}$, $Q_{\mathrm{t}}$, $r_\mathrm{b}$ and $r_\mathrm{t}$ are the nuisance parameters of the method and are assumed to be Gaussian random variables.
The Gaussian nature of the combined charges was tested with 10 million MC experiments. In each experiment the mean combined charge was found by averaging 1000 combined charges generated from a MC-simulated combined charge spectrum for the  muon channel. The obtained distribution of the mean combined charges was normally distributed and the Gaussian fit to the distribution showed a goodness-of-fit of  $\chi^2/ndf $ = 86/103.    
Their uncertainties scaled to the data integrated luminosity (2.05 \ifb) are summarized in table~\ref{nuisance}.
\begin{table*}[htb]
 \begin{center}
 \begin{tabular}{c|c|c|c|c|c}
 \hline \hline
 Channel   	& $Q_{\mathrm{s}}$    & $Q_{\mathrm{b}}$   &  $Q_{\mathrm{t}}$    & $r\mathrm{_{b}}$    & $r\mathrm{_{t}}$   \\ \hline
 $e$       	& -0.080 $\pm$ 0.007  & -0.015 $\pm$ 0.041 &  -0.066 $\pm$ 0.042  &  0.066 $\pm$ 0.018  & 0.042 $\pm$ 0.012 \\
 $\mu$   	& -0.078 $\pm$ 0.006  & -0.052 $\pm$ 0.028 &  -0.051 $\pm$ 0.038  &  0.088 $\pm$ 0.025  & 0.038 $\pm$ 0.011 \\
 $e+\mu$   	& -0.079 $\pm$ 0.004  & -0.038 $\pm$ 0.023 &  -0.058 $\pm$ 0.028  &  0.079 $\pm$ 0.016  & 0.040 $\pm$ 0.008 \\
\hline \hline
\end{tabular}
 \bigskip
 \caption{The nuisance parameters: the expected combined charge mean values and their standard deviations  for the  signal ($Q_{\mathrm{s}}$), non-top-quark background ($Q_{\mathrm{b}}$), single-top-quark background ($Q_{\mathrm{t}}$) and the fractions of non-top-quark ($r_\mathrm{b}$) and single-top-quark ($r_\mathrm{t}$)  backgrounds for an integrated luminosity of 2.05 \ifb.}
\label{nuisance}
\end{center}
\end{table*}
\begin{figure}[ht!]
\begin{center}
\begin{tabular}{cc}
\mbox{\includegraphics[height=6cm]{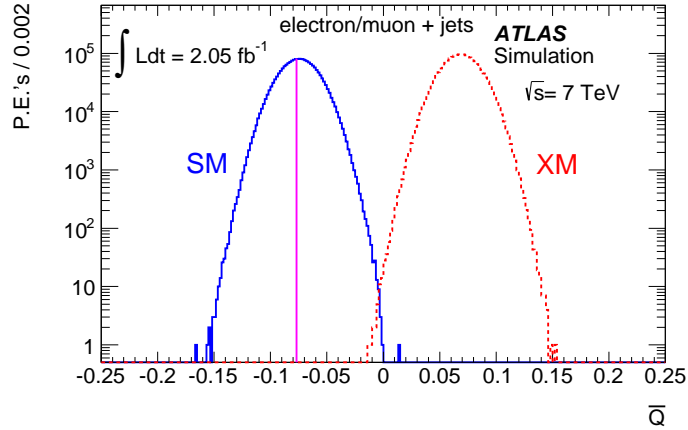}}
\end{tabular}
\end{center}
\caption{The expected distribution of the mean value of the combined charge, $\bar Q$, for the electron and muon channels resulting from  pseudo-experiments for the SM (solid blue line) and the exotic (dashed red line) hypothesis for an integrated luminosity of 2.05~\ifb. The magenta vertical line represents the value measured in the data.} 
\label{fig:SMvsXM}
\end{figure}
The two hypotheses are compared by calculating the $p$-value \citep{james_2006}, the probability of obtaining a test statistic at least as extreme as
the one that was actually observed provided that the null hypothesis is true. In order to obtain the $p$-value  for the observed values of the test statistic $\langle Q_{\mathrm{comb}}\rangle$
(see the data column of table~\ref{data_Qcomb}),
pseudo-experiments for both hypotheses, the SM as well as the exotic model, have been performed.
To take into account a possible difference between MC and experimental data, a scale factor ($\mathrm{SF}$) is defined as the ratio of
experimental to MC mean combined  charges for a QCD $b$-jet sample.
The scale factor $\mathrm{SF}$ was found using double  $b$-tagged dijet events containing a soft muon, where the charge of the soft muon determines the flavour of the $b$-jet (i.e. if $b$ or $\bar{b}$ initiated the jet).  This technique gives $\mathrm{SF} = 1.00$ with a spread $\sigma = 0.19$. The technique based on the absolute value of the $b$-jet charge, i.e. based on the data-to-MC ratio from figure~\ref{fig:bQ_pT}, leads to a scale factor compatible with unity with a spread $\sigma = 0.02$. To be conservative, the former value is used. 
The $\mathrm{SF}$ uncertainty is added in quadrature to the statistical and systematic uncertainties of the combined mean charge.

In figure~\ref{fig:SMvsXM} the distributions from the pseudo-experiments of the observed mean combined charge ($\bar Q$) are shown for both hypotheses, the SM (solid blue line) and the exotic model (dashed red line). The magenta line in this plot corresponds to the experimentally observed value $Q_\mathrm{obs}$. The figure shows the results for the combined electron and muon channels.
Each of these distributions is obtained from pseudo-experiments in which the nuisance parameters are sampled from  Gaussian distributions with the mean values and standard deviations taken
from table~\ref{nuisance}. In addition, the sampled charge $\bar Q$ is Gaussian-smeared by the mean combined charge systematic uncertainty and by the SF uncertainty.

 The $p$-values for the SM and the exotic model,
the distance of $Q_\mathrm{obs}$ from the expected value of the exotic combined charge in standard deviations, and the quantities $\alpha$ and $ \beta$, are summarized in table~\ref{pvalue} for the combined electron and muon ($e+\mu $) channel as well as for the electrons ($e$) and muons ($\mu $) channels separately. 
\begin{table*}[htb]
 \begin{center}
 \begin{tabular}{c|c|c|c|c|c}
 \hline \hline
  Channel  & $p_\mathrm{SM}$ & $p_\mathrm{XM}$  & $\sigma_\mathrm{XM}$(s.d.)& $\alpha$   & $\beta$             \\ \hline
  $e$      & 0.813            & $< 10^{-7}$    &            8.8            & $6.1\times 10^{-6}$ & $1.1\times 10^{-5}$    \\
  $\mu $   & 0.960            & $< 10^{-7}$    &            8.5            & $2.4\times 10^{-6}$ & $4.0\times 10^{-5}$    \\
  $e+\mu $ & 0.892            & $< 10^{-7}$    &            8.9            & $2.4\times 10^{-6}$ & $1.5\times 10^{-5}$    \\
\hline
\hline

\end{tabular}
 \bigskip
 \caption{The $p$-values for the SM ($p_\mathrm{SM}$) and exotic model ($p_\mathrm{XM}$); the distance $\sigma_\mathrm{XM}$ of the observed value, $Q_\mathrm{obs}$, from the expected value of the exotic combined charge in standard deviations (s.d.); the significance level ($\alpha $) and the false negative rate ($\beta $) for the integrated luminosity of 2.05 \ifb. }
\label{pvalue}
\end{center}
\end{table*}

From table~\ref{pvalue} it can be seen that the data are fully compatible with the SM. The $p$-values for the SM scenario are high (the two-sided $p$-value is more than 80\%) while 
those for the exotic hypothesis are very small (less than 10$^{-7}$). 
None of the 20 million exotic-hypothesis pseudo-experiments have $\bar Q$ values below the observed value of the mean combined charge.
Converting the $p$-value into the number of standard deviations for
the exotic-scenario mean combined charge distribution, an exclusion at the level higher than 8$\sigma $ is obtained for the combination of the electron and muon channels. This result assumes Gaussian-distributed nuisance parameters, as  supported by the performed MC experiments. 
Due to fact that most of the systematic uncertainties were combined and are common to the electron and muon channels, the differences in the nuisance parameters do not lead to large differences in the exclusion limits for the individual channels.  

\section{Conclusion}
\begin{sloppypar}
\indent The top quark charge has been studied using 2.05~fb$^{-1}$ of data accumulated  by the ATLAS experiment at a centre-of-mass energy of 7 TeV.
The measured top quark charge is \mbox{0.64~$\pm $~0.02~(stat.)~$\pm $~0.08 (syst.)}. This result strongly favours the \SM\ and excludes  models  with an exotic quark with charge --4/3 instead of the top quark by more than 8$\sigma $.
\end{sloppypar}

\section{Acknowledgements}
We thank CERN for the very successful operation of the LHC, as well as the
support staff from our institutions without whom ATLAS could not be
operated efficiently.

We acknowledge the support of ANPCyT, Argentina; YerPhI, Armenia; ARC,
Australia; BMWF and FWF, Austria; ANAS, Azerbaijan; SSTC, Belarus; CNPq and FAPESP,
Brazil; NSERC, NRC and CFI, Canada; CERN; CONICYT, Chile; CAS, MOST and NSFC,
China; COLCIENCIAS, Colombia; MSMT CR, MPO CR and VSC CR, Czech Republic;
DNRF, DNSRC and Lundbeck Foundation, Denmark; EPLANET, ERC and NSRF, European Union;
IN2P3-CNRS, CEA-DSM/IRFU, France; GNSF, Georgia; BMBF, DFG, HGF, MPG and AvH
Foundation, Germany; GSRT and NSRF, Greece; ISF, MINERVA, GIF, DIP and Benoziyo Center,
Israel; INFN, Italy; MEXT and JSPS, Japan; CNRST, Morocco; FOM and NWO,
Netherlands; BRF and RCN, Norway; MNiSW, Poland; GRICES and FCT, Portugal; MERYS
(MECTS), Romania; MES of Russia and ROSATOM, Russian Federation; JINR; MSTD,
Serbia; MSSR, Slovakia; ARRS and MIZ\v{S}, Slovenia; DST/NRF, South Africa;
MICINN, Spain; SRC and Wallenberg Foundation, Sweden; SER, SNSF and Cantons of
Bern and Geneva, Switzerland; NSC, Taiwan; TAEK, Turkey; STFC, the Royal
Society and Leverhulme Trust, United Kingdom; DOE and NSF, United States of
America.

The crucial computing support from all WLCG partners is acknowledged
gratefully, in particular from CERN and the ATLAS Tier-1 facilities at
TRIUMF (Canada), NDGF (Denmark, Norway, Sweden), CC-IN2P3 (France),
KIT/GridKA (Germany), INFN-CNAF (Italy), NL-T1 (Netherlands), PIC (Spain),
ASGC (Taiwan), RAL (UK) and BNL (USA) and in the Tier-2 facilities
worldwide.

\newpage
\bibliographystyle{JHEP}
\bibliography{tc}

\providecommand{\href}[2]{#2}\begingroup\raggedright\begin{thebibliography}{10}

\bibitem{abe95}
{\bf CDF} Collaboration, F.~Abe et~al., {\it {Observation of Top Quark
  Production in $\overline{\mathit{p}}\mathit{p}$ Collisions with the Collider
  Detector at Fermilab}},  {\em Phys. Rev. Lett.} {\bf 74} (1995) 2626.

\bibitem{d095}
{\bf D0} Collaboration, S.~Abachi et~al., {\it {Observation of the Top Quark}},
   {\em Phys. Rev. Lett.} {\bf 74} (1995) 2632.

\bibitem{chang00}
D.~Chang, W.-F. Chang, and E.~Ma, {\it {Alternative interpretation of the
  Fermilab Tevatron top events}},  {\em Phys. Rev. {\bf D}} {\bf 59} (1999)
  091503.

\bibitem{D0_topQ}
{\bf D0} Collaboration, V.~M. Abazov et~al., {\it {Experimental Discrimination
  between Charge $2e/3$ Top Quark and Charge $4e/3$ Exotic Quark Production
  Scenarios}},  {\em Phys. Rev. Lett.} {\bf 98} (2007) 041801.

\bibitem{CDF_topQ_2}
{\bf CDF} Collaboration, T.~Aaltonen et~al., {\it {Exclusion of an Exotic Top
  Quark with 4/3 Electric Charge Using Soft Lepton Tagging}},  {\em Phys. Rev.
  Lett.} {\bf 105} (2010) 101801.

\bibitem{cdfnew}
{\bf CDF} Collaboration, {\it {Exclusion of exotic top-like quarks with -4/3
  electric charge using jet-charge tagging in single-lepton ttbar events at
  CDF}},  {\em {submitted to Phys. Rev. D}} (2013)
  [\href{http://xxx.lanl.gov/abs/1304.4141}{{\tt arXiv:1304.4141}}].

\bibitem{AtlDet2}
{\bf ATLAS} Collaboration, {\it {The ATLAS Experiment at the CERN Large Hadron
  Collider}},  {\em JINST} {\bf 3} (2008) S08003.

\bibitem{Aad:2013lumi}
{\bf ATLAS} Collaboration, {\it Improved luminosity determination in pp
  collisions at $\sqrt{s}=$ 7~tev using the atlas detector at the lhc},  {\em
  submitted to Eur. Phys. J. C} (2013)
  [\href{http://xxx.lanl.gov/abs/1302.4393}{{\tt arXiv:1302.4393}}].

\bibitem{Geant4_2003}
{\bf GEANT4} Collaboration, S.~Agostinelli et~al., {\it {GEANT4: A Simulation
  toolkit}},  {\em Nucl. Instrum. Meth.} {\bf A 506} (2003) 250.

\bibitem{atlas_sim}
{\bf ATLAS} Collaboration, {\it {The ATLAS Simulation Infrastructure}},  {\em
  Eur. Phys. J.} {\bf C 70} (2010) 823.

\bibitem{MCatNLO}
S.~Frixione and B.~R. Webber, {\it {Matching NLO QCD computations and parton
  shower simulations}},  {\em JHEP} {\bf 06} (2002) 029,
  [\href{http://xxx.lanl.gov/abs/hep-ph/0204244}{{\tt hep-ph/0204244}}].

\bibitem{Frixione:2003ei}
S.~Frixione, P.~Nason, and B.~R. Webber, {\it {Matching NLO QCD and parton
  showers in heavy flavour production}},  {\em JHEP} {\bf 08} (2003) 007,
  [\href{http://xxx.lanl.gov/abs/hep-ph/0305252}{{\tt hep-ph/0305252}}].

\bibitem{PhysRevD.78.013004}
P.~M. Nadolsky et~al., {\it {Implications of CTEQ global analysis for collider
  observables}},  {\em Phys. Rev. {\bf D}} {\bf 78} (2008) 013004.

\bibitem{Herwig}
G.~Corcella et~al., {\it {HERWIG 6: An event generator for hadron emission
  reactions with interfering gluons (including supersymmetric processes)}},
  {\em JHEP} {\bf 0101} (2001) 010,
  [\href{http://xxx.lanl.gov/abs/hep-ph/0011363}{{\tt hep-ph/0011363}}].

\bibitem{Frixione:2010ra}
S.~Frixione, F.~Stoeckli, P.~Torrielli, and B.~R. Webber, {\it {NLO QCD
  corrections in Herwig++ with MC@NLO}},  {\em JHEP} {\bf 01} (2011) 053,
  [\href{http://xxx.lanl.gov/abs/1010.0568}{{\tt arXiv:1010.0568}}].

\bibitem{jimmy}
J.~Butterworth, J.~Forshaw, and M.~Seymour, {\it {Multiparton interactions in
  photoproduction at HERA}},  {\em {Z. Phys.}} {\bf C 72} (1996) 637.

\bibitem{powheg}
S.~Frixione, P.~Nason, and C.~Oleari, {\it {Matching NLO QCD computations with
  parton shower simulations: the POWHEG method}},  {\em JHEP} {\bf 11} (2007)
  070, [\href{http://xxx.lanl.gov/abs/hep-ph/0409146}{{\tt hep-ph/0409146}}].

\bibitem{Pythia}
T.~Sjostrand, S.~Mrenna, and P.~Skands, {\it {PYTHIA 6.4 Physics and Manual}},
  {\em JHEP} {\bf 0605} (2006) 026,
  [\href{http://xxx.lanl.gov/abs/hep-ph/0603175}{{\tt hep-ph/0603175}}].

\bibitem{ACERMC}
B.~Kersevan and E.~Richter-Was, {\it {The Monte Carlo Event Generator AcerMC
  versions 2.0 to 3.8 with interfaces to PYTHIA 6.4, HERWIG 6.5 and ARIADNE
  4.1}},  \href{http://xxx.lanl.gov/abs/hep-ph/0405247}{{\tt hep-ph/0405247}}.

\bibitem{hathor_10}
M.~Aliev et~al., {\it {Hadronic Top and Heavy Quarks Cross Section
  Calculator}},  {\em Comput. Phys. Commun.} {\bf 182} (2011) 1034,
  [\href{http://xxx.lanl.gov/abs/1007.1327}{{\tt arXiv:1007.1327}}].

\bibitem{kidonakis_D83}
N.~Kidonakis, {\it {Next-to-next-to-leading-order collinear and soft gluon
  corrections for t-channel single top quark production}},  {\em Phys. Rev.
  {\bf D}} {\bf 83} (2011) 091503.

\bibitem{kidonakis_D81}
N.~Kidonakis, {\it {Next-to-next-to-leading logarithm resummation for s-channel
  single top quark production}},  {\em Phys. Rev. {\bf D}} {\bf 81} (2010)
  054028.

\bibitem{kidonakis_D82}
N.~Kidonakis, {\it {Two-loop soft anomalous dimensions for single top quark
  associated production with a $W^-$ or $H^-$}},  {\em Phys. Rev. {\bf D}} {\bf
  82} (2010) 054018.

\bibitem{Alpgen}
M.~Mangano et~al., {\it {ALPGEN, a generator for hard multiparton processes in
  hadronic collisions}},  {\em JHEP} {\bf 0307} (2003) 001,
  [\href{http://xxx.lanl.gov/abs/hep-ex/0206293}{{\tt hep-ex/0206293}}].

\bibitem{Pumplin:2002vw}
J.~Pumplin et~al., {\it {New generation of parton distributions with
  uncertainties from global QCD analysis}},  {\em JHEP} {\bf 07} (2002) 012,
  [\href{http://xxx.lanl.gov/abs/hep-ph/0201195}{{\tt hep-ph/0201195}}].

\bibitem{melnikov_06}
K.~Melnikov and F.~Petriello {\em Phys. Rev. {\bf D}} {\bf 74} (2006) 114017,
  [\href{http://xxx.lanl.gov/abs/hep-ph/0609070}{{\tt hep-ph/0609070}}].

\bibitem{mrst2007lo}
A.~Sherstnev and R.~Thorne, {\it {Parton distributions for LO generators}},
  {\em Eur. Phys. J. \bf C} {\bf 55} (2008) 553.

\bibitem{PythiaAMBT2B}
{ATLAS Collaboration}, {\it {Further tunes of PYTHIA6 and Pythia 8}},
  ATL-PHYS-PUB-2011-014.

\bibitem{electronCP}
{\bf ATLAS} Collaboration, {\it {Electron performance measurements with the
  ATLAS detector using the 2010 LHC proton-proton collision data}},  {\em Eur.
  Phys. J.} {\bf C 72} (2012) 1909,
  [\href{http://xxx.lanl.gov/abs/1110.3174}{{\tt arXiv:1110.3174}}].

\bibitem{ATLAS-CONF-2011-063}
{\bf ATLAS} Collaboration, {\it {Muon reconstruction efficiency in reprocessed
  2010 LHC proton-proton collision data recorded with the ATLAS detector}},
  ATLAS-CONF-2011-063.

\bibitem{Cacciari_2008}
M.~Cacciari, G.~P. Salam, and G.~Soyez, {\it {The anti-$k_t$ jet clustering
  algorithm}},  {\em JHEP,} {\bf 04} (2008) 63.

\bibitem{jet_syst}
{\bf ATLAS} Collaboration, {\it {Jet energy measurement with the ATLAS detector
  in proton-proton collisions at $\sqrt{s} =$~7~\TeV}},  {\em Eur. Phys. J.}
  {\bf C 73} (2011) 2304, [\href{http://xxx.lanl.gov/abs/1112.6426}{{\tt
  arXiv:1112.6426}}].

\bibitem{topoCluster}
W.~Lampl et~al., {\it {Calorimeter Clustering Algorithms: Description and
  Performance}},
  \href{http://xxx.lanl.gov/abs/http://cds.cern.ch/record/1099735}{{\tt
  http://cds.cern.ch/record/1099735}}. ATL-LARG-PUB-2008-002.

\bibitem{metCP}
{\bf ATLAS} Collaboration, {\it {Performance of missing transverse momentum
  reconstruction in proton-proton collisions at $\sqrt{s}$ = 7~TeV with
  ATLAS}},  {\em Eur. Phys. J. \bf C} {\bf 72} (2012) 1844.

\bibitem{JetFitterCombNN}
{\bf ATLAS} Collaboration, {\it {Measuring the b-tag efficiency in a top-pair
  sample with 4.7 fb$^{-1}$ of data from the ATLAS detector}},
  \href{http://xxx.lanl.gov/abs/http://cdsweb.cern.ch/record/1460443}{{\tt
  http://cdsweb.cern.ch/record/1460443}}. ATLAS-CONF-2012-097.

\bibitem{JetFitterTag}
{\bf ATLAS} Collaboration, {\it {Commissioning of the ATLAS high-performance
  b-tagging algorithms in the 7~TeV collision data}},
  \href{http://xxx.lanl.gov/abs/http://cdsweb.cern.ch/record/1369219}{{\tt
  http://cdsweb.cern.ch/record/1369219}}. ATLAS-CONF-2011-102.

\bibitem{ttbarXs_2011}
{\bf ATLAS} Collaboration, {\it {Measurement of the top quark-pair production
  cross section with ATLAS in pp collisions at $\sqrt{s} $= 7~TeV}},  {\em Eur.
  Phys. J. \bf C} {\bf 71} (2011) 1577,
  [\href{http://xxx.lanl.gov/abs/hep-ex/1012.1792}{{\tt hep-ex/1012.1792}}].

\bibitem{Field_Feynman}
R.~Field and R.~Feynman, {\it {A parameterization of the properties of quark
  jets}},  {\em Nucl. Phys.} {\bf B 136} (1978) 1.

\bibitem{Aleph_b-charge}
{\bf ALEPH} Collaboration, R.~Barate et~al., {\it {Determination of A(b)(FB)
  using jet charge measurements in Z decays}},  {\em Phys. Lett.} {\bf B 426}
  (1998) 217.

\bibitem{Aad2012418}
{\bf ATLAS} Collaboration, {\it {Measurement of the cross section for the
  production of a W boson in association with b-jets in pp collisions at
  $\sqrt{s}$ = 7~TeV with the ATLAS detector}},  {\em Phys. Lett. \bf B} {\bf
  707} (2012) 418.

\bibitem{ttbarDifXs_2013}
{\bf ATLAS} Collaboration, {\it {Measurement of top quark pair relative
  differential cross-sections with ATLAS in pp collisions at $\sqrt{s}=$
  7~TeV}},  {\em Eur. Phys. J. \bf C} {\bf 73} (2013) 2261,
  [\href{http://xxx.lanl.gov/abs/hep/ph.1207.5644}{{\tt hep/ph.1207.5644}}].

\bibitem{MC_Perugia}
P.~Skands, {\it {Tuning Monte Carlo Generators: The Perugia Tunes}},  {\em
  Phys. Rev. {\bf D}} {\bf 82} (2010) 074018,
  [\href{http://xxx.lanl.gov/abs/arxiv:1005.3457v4}{{\tt arxiv:1005.3457v4}}].

\bibitem{color_rec1}
{\bf The TeV4LHC QCD Working Group} Collaboration, M.~Albrow et~al., {\it
  {Tevatron-for-LHC report of the QCD Working Group}},
  \href{http://xxx.lanl.gov/abs/hep-ph/0610012}{{\tt hep-ph/0610012}}.

\bibitem{single_top_syst}
N.~Kidonakis, {\it {Single top quark production cross section at hadron
  colliders}},  {\em PoS} {\bf DIS2010} (2010) 196,
  [\href{http://xxx.lanl.gov/abs/1005.3330}{{\tt arXiv:1005.3330}}].

\bibitem{jet_scale_EPJC_11}
{\bf ATLAS} Collaboration, {\it {Measurement of inclusive jet and dijet cross
  sections in proton-proton collisions at 7 TeV centre-of-mass energy with the
  ATLAS detector}},  {\em Eur. Phys. J. \bf C} {\bf 71} (2011) 1512,
  [\href{http://xxx.lanl.gov/abs/1009.5908}{{\tt arXiv:1009.5908}}].

\bibitem{jet_scale_PRD_12}
{\bf ATLAS} Collaboration, {\it {Measurement of inclusive jet and dijet
  production in pp collisions at $\sqrt {s}$~=7~TeV using the ATLAS detector}},
   {\em Phys. Rev. {\bf D}} {\bf 86} (2012) 014022.

\bibitem{PDG2012}
{\bf {Particle Data Group}} Collaboration, J.~Beringer et~al., {\it {2012
  Review of Particle Physics}},  {\em {Phys. Rev. D}} {\bf 86} (2012) 010001.

\bibitem{CandH_1992}
R.~D. Cousins and V.~L. Highland, {\it {Incorporating systematic uncertainties
  into an upper limit}},  {\em Nucl. Instrum. Meth.} {\bf A 320} (1992) 331.

\bibitem{james_2006}
F.~James, {\em {Statistical Methods in Experimental Physics}}.
\newblock North-Holland Publishing Co., {(2006)}.

\bibitem{cowan_1998}
G.~Cowan, {\em {Statistical Data Analysis}}.
\newblock {Clarendon Press}, {(1998)}.

\end{thebibliography}\endgroup

\onecolumn
\clearpage
\begin{flushleft}
{\Large The ATLAS Collaboration}

\bigskip

G.~Aad$^{\rm 48}$,
T.~Abajyan$^{\rm 21}$,
B.~Abbott$^{\rm 112}$,
J.~Abdallah$^{\rm 12}$,
S.~Abdel~Khalek$^{\rm 116}$,
A.A.~Abdelalim$^{\rm 49}$,
O.~Abdinov$^{\rm 11}$,
R.~Aben$^{\rm 106}$,
B.~Abi$^{\rm 113}$,
M.~Abolins$^{\rm 89}$
% ...
\bigskip
\\
% ...
$^{11}$ Institute of Physics, Azerbaijan Academy of Sciences, Baku, Azerbaijan\\
$^{12}$ Institut de F{\'\i}sica d'Altes Energies and Departament de F{\'\i}sica de la Universitat Aut{\`o}noma de Barcelona and ICREA, Barcelona, Spain\\
% ...
$^{21}$ Physikalisches Institut, University of Bonn, Bonn, Germany\\
% ...
$^{48}$ Fakult{\"a}t f{\"u}r Mathematik und Physik, Albert-Ludwigs-Universit{\"a}t, Freiburg, Germany\\
$^{49}$ Section de Physique, Universit{\'e} de Gen{\`e}ve, Geneva, Switzerland\\
% ...
$^{89}$ Department of Physics and Astronomy, Michigan State University, East Lansing MI, United States of America\\
% ...
$^{106}$ Nikhef National Institute for Subatomic Physics and University of Amsterdam, Amsterdam, Netherlands\\
% ...
$^{112}$ Homer L. Dodge Department of Physics and Astronomy, University of Oklahoma, Norman OK, United States of America\\
$^{113}$ Department of Physics, Oklahoma State University, Stillwater OK, United States of America\\
% ...
$^{116}$ LAL, Universit{\'e} Paris-Sud and CNRS/IN2P3, Orsay, France\\
% ...
\end{flushleft}


\end{document}